\documentclass[twocolumn,amsmath,amssymb,prl]{revtex4-1}
\usepackage{physics}
\usepackage[dvipdfmx]{graphicx} 
\usepackage{dcolumn} % Align table columns on a decimal point
\usepackage{bm} % bold ma
\usepackage{braket} % braket
\usepackage{ulem} % sout
\usepackage[usenames,dvipsnames]{color}% use as \textcolor{red}{******}
\usepackage[%dviepsmx,
colorlinks=true, citecolor=blue, urlcolor=blue, linkcolor=blue,
setpagesize=false, bookmarks=false
]{hyperref}
\usepackage{siunitx}
\usepackage{mhchem}
\usepackage{comment}
\usepackage{mathrsfs}

\begin{document}

%%%%%%%%%%%%%%%%%%%%
% Title and authors
%%%%%%%%%%%%%%%%%%%%

\title{%THz-pulse driven 
Triplons, triplon pairs and dynamical symmetries in laser-driven Shastry-Sutherland magnets}
\author{Mina Udono}
\author{Masahiro Sato}
\affiliation{Department of Physics, Chiba University, Chiba 263-8522, Japan}
\date{\today}

%%%%%%%%%%%%%%%%%%%%
% Abstract
%%%%%%%%%%%%%%%%%%%%

\begin{abstract}
In frustrated magnets, a variety of novel phenomena arise due to their peculiar magnetic states, whose excitations usually reside in the GHz to THz range. 
%and these magnets have been explored for a long time. 
Intense THz lasers, therefore, lead to examining nonlinear optical effects in such magnets. 
Employing an unbiased numerical method, we calculate the laser-pulse driven harmonic spectra of the Shastry-Sutherland magnets, which are well known as the model of a quantum frustrated magnet~\ce{SrCu2(BO3)2}. 
As a result, nonlinear responses of triplons and a 
two-triplon bound state are numerically shown to emerge
through the Zeeman or magnetoelectric couplings between the laser and the electron spin.
%depending on the observed physical quantities. 
Furthermore, the dynamical symmetries and selection rules of harmonic spectra for the Shastry-Sutherland model, realized in ~\ce{SrCu2(BO3)2}, are derived.
The selection rules tell us that one can obtain information on magnetic anisotropy and symmetry from the spectra by manipulating external fields or two-color lasers. 
Our results indicate that laser-driven harmonic spectra are useful 
as a probe of %invisible 
less-detectable quantum many-body states in frustrated magnets. 
%for understanding natures of quantum states in frustrated magnets. 
\end{abstract}

\maketitle

%%%%%%%%%%%%%%%%%%%%
% Introduction
%%%%%%%%%%%%%%%%%%%%

%\section{Introduction}
{\it Introduction}-- 
Quantum frustrated magnets are a treasure trove of exotic physical phenomena due to the prominent quantum fluctuations and many-body effects. 
The two-dimensional (2D) spin-$\frac{1}{2}$ Shastry-Sutherland model (SSM)~\cite{Shastry1981}, which consists of two types of orthogonal spin dimers, is a typical frustrated magnetic model. The SSM not only shows %various 
interesting physical properties due to the peculiar structure but also approximately describes the magnetic properties of \ce{SrCu2(BO3)2} (SCBO)~\cite{Miyahara2003,Knetter2000,Totsuka2001,Cepas2001,Kageyama2002,Knetter2004,Romhanyi2011,Corboz2014,Wietek2019,Jimenez2021,Shi2022}, which Kageyama synthesized in 1999~\cite{Kageyama1999}. 
Various phenomena have been discovered in SCBO~\cite{Takigawa2010,Kageyama1999-2,Nojiri1999,Kageyama2000,Kageyama2000-2,Lemmens2000,Nojiri2003,Gaulin2004,Zorko2004,Kakurai2005,Tsujii2011,Bettler2020,Imajo2022}.

\begin{figure}[!h]
\begin{center}
\includegraphics[width=\columnwidth]{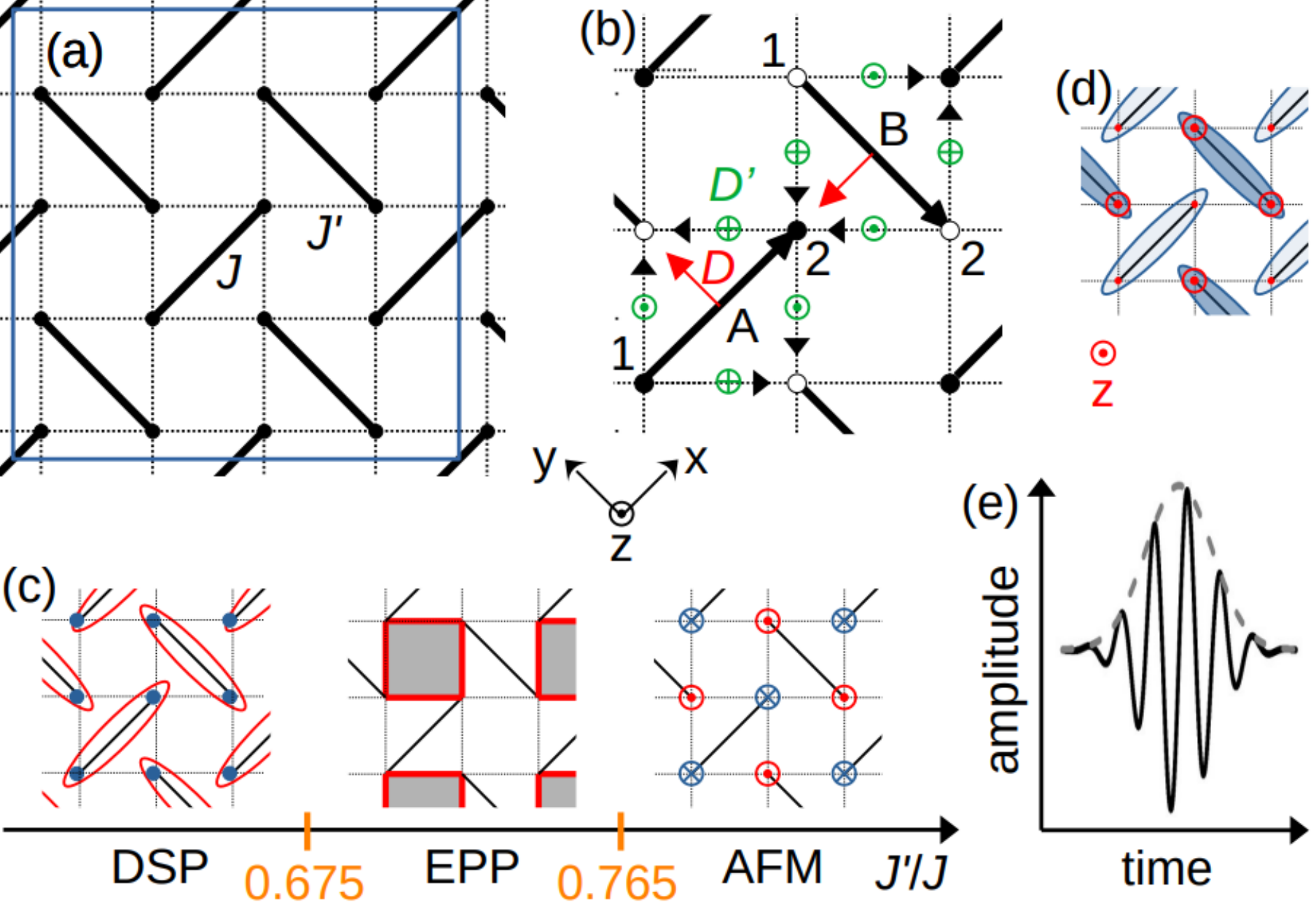}
\caption{
(a) Lattice of SSM. 
Thick and thin dashed lines, respectively, indicate the intradimer and interdimer antiferromagnetic interactions. 
Our calculation employs the $16$-site square cluster, surrounded by the blue line. 
(b) Symmetry allowed intradimer (red) and interdimer (green) DM vectors in a unit cell. 
Each black arrow from i to j shows the ordering of two spin operators $\hat{\bm{S}}_i\cross\hat{\bm{S}}_j$ in the DM term on the i-j bond. 
Orthogonal dimers (thick continuous lines) are denoted by A (consists of filled circle sites) and B (consists of open circle sites); spin sites on the dimer are numbered 1 and 2. 
The $\hat{S}^\alpha$ axis is parallel to the real-space $\alpha$ axis depicted below the panel (b). 
(c) Ground-state phase diagram of the SSM. 
The DSP, EPP, and AFM refer to the dimer singlet phase, the empty plaquette phase, and the antiferromagnetic phase, respectively~\cite{Corboz2013}. 
%Phase transition points obtained by iPEPS~\cite{Corboz2013} are filled in. 
(d) Spin configuration of the 1/2 magnetization plateau phase (PP). Two distinct dimers exist; dimers with $\langle \hat{S}^z_j\rangle=1/4+\delta$ (red double circles) and ones with $\langle \hat{S}^z_j\rangle=1/4-\delta$ (small red circles). 
%and small magnetization ($\langle \hat{S}^z_j\rangle=1/4-\delta$) denoted by red circles.
%Filled circles, whose sizes are proportional to the spin amplitude, show sites with magnetization along the magnetic field. 
(e) Schematic of Gaussian pulse used in our simulations. 
The dashed line indicates the Gaussian envelope. 
}
\label{fig:model_SS}
\end{center}
\end{figure}

The SSM possesses an intradimer interaction $J(>0)$ and interdimer one $J'(>0)$, 
and its lattice and unit cell are shown in Fig.~\ref{fig:model_SS}(a) and (b), respectively. 
The ground state of the SSM has at least two phases depending on the ratio of $J$ to $J'$: 
%and the quantum phase transition occurs. 
One is a dimer singlet phase (DSP) with a spin gap in the $J\gg J'$ limit and the other is a gapless antiferromagnetic (i.e., N\'eel) ordered state in the $J\ll J'$ limit. 
The intermediate phase between these two phases is considered to be the plaquette state with four spins forming a singlet~\cite{Koga2000,Takushima2001,Lauchli2002}. %, the transition to this phase occurs under pressure. 
The ground-state phase diagram of the SSM is depicted in Fig.~\ref{fig:model_SS}(c)~\cite{Koga2000,Corboz2013,Xi2023}. 
The SCBO with $J'/J\simeq 0.635$, whose ground state is in the DSP~\cite{Miyahara1999,Kageyama1999}, is close to the quantum critical point of $J'/J\simeq 0.67$ between the dimer singlet and plaquette phases. 
A plaquette phase has indeed been realized by applying a huge pressure %(GPa order) 
to SCBO~\cite{Sakurai2009,Waki2007,Haravifard2012,Sakurai2018,Zayed2017,Guo2020,Yi2023} and its properties have been actively explored. 
In addition to pressure, an external magnetic field is generally useful to generate novel phases in frustrated magnets. In fact, several magnetization plateau phases have been observed and investigated in SCBO~\cite{Kageyama1999,Onizuka2000,Kodama2002,Momoi2000-2,Miyahara2000-2,Momoi2000,Takigawa2004, Kodama2005,Suchitra2008,Marcelo2012,Takigawa2013,Matsuda2013,Haravifard2016,Nomura2023}. 

Many experimental techniques are available to observe exotic properties of frustrated magnets including SCBO. 
For instance, measurements of magnetization, susceptibilities, and specific heat are very powerful to see static properties. 
Nuclear magnetic resonance, electron spin resonance, neutron scattering, thermal transport and Raman scattering are useful to investigate dynamical natures or excitations in magnets. 

Besides these established methods, THz-laser techniques (electromagnetic waves in $10^{11-13}$ Hz range) have been developed in the last decades~\cite{Hirori2011,Sato2013,Hafez2016,Dhillon2017,Liu2017-2,Mukai2016,Kovalev2018,Evangelos2021,Alfred2023}. 
THz-photon energies (0.1-10 meV) are comparable to those of magnetic excitations. Hence, THz waves are very useful to directly observe such excitations~\cite{Nicoletti2016,Kampfrath2011,Kampfrath2013,Vicario2013,Baierl2016-2,Mukai2016,Lu2017-2,Bossini2019,Li2020,Walowski2016,Salen2019,Safin2020,Pimenov2006,Takahashi2012,Kubackaand2014,Wang2018}. In fact, various magnetic quasiparticles like magnons~\cite{Kampfrath2011,Kampfrath2013,Vicario2013,Baierl2016-2,Mukai2016,Lu2017-2,Bossini2019,Li2020,Walowski2016,Salen2019,Safin2020}, electromagnons~\cite{Pimenov2006,Takahashi2012,Kubackaand2014}, magnetic solitons~\cite{Wang2018}, etc. have been detected with THz-wave methods. Moreover, quite recently, the harmonic generation (HG) up to the third harmonic via magnetic excitations (without any photo-induced electron transition) has been observed by applying intense \ce{THz} pulses to an antiferromagnet~\cite{Zhang2023}. 
%Motivated by these experimental developments, 
In addition to experiments,
THz-wave induced phenomena have been theoretically predicted: %and proposed: 
Control of magnetic states with intense THz wave~\cite{Mochizuki2010-2}, 
THz-laser-driven nonlinear responses in quantum spin chains~\cite{Ikeda2019, Allafi2024}, spin-nematic magnets~\cite{Sato2020}, and Kitaev magnets~\cite{Kanega2021}, and Floquet engineering of magnets~\cite{Sato2014,Sato2016,Higashikawa2018,Hirosawa2022,Yambe2023,Yambe2024,Shin2018}.

In this paper, motivated by these %research 
activities, we theoretically explore the THz-laser-driven nonlinear optical response of the SSM with/without some perturbations. 
From the theoretical viewpoint, numerical studies of non-equilibrium many-body states in 2D models are challenging 
because it is generally difficult to analyze sufficiently large 2D systems due to computational cost. 
However, by using an unbiased numerical method for finite-size systems, 
%based on the exact diagonalization, 
we obtain reliable results about the THz-pulse-induced spin dynamics in the SSMs. We mainly focus on the DSP 
%dimer-singlet phase 
realized in SCBO, where triplons with spin $S=1$ and two-triplon bound states (triplon pairs) appear as the low-energy excitations~\cite{Momoi2000,Kageyama2000,Totsuka2001,Nojiri2003,Gozar2005,Romhanyi2011,McClarty2017,Wulferding2021,Fogh2024}. 
We show that triplons or a triplon pair are excited by a THz pulse, and %as a result, 
first- to third-order harmonics are observed depending on the sorts of light-matter couplings (Zeeman and magnetoelectric couplings) and the light polarization direction. 
Namely, the dynamical symmetries~\cite{Alon1998,Neufeld2019,Ikeda2019,Ikeda2020,Kanega2021,Kanega2024,Neufeld2023} associated with the light-matter coupling are shown to drastically change the selection rule (appearance or absence) of the $n$-th order harmonics.
%{\color{red} Moreover, selection rules related to dynamical symmetries were derived for time dependent spin responses as well as for the total electronic excitation, which are important factors in the magnetization dynamics~\cite{Neufeld2023}.}
We find that the selection rule quite differs from those in photo-driven electron systems accompanying electric dipole transition, and it can be controlled by introducing static perturbations (like external magnetic fields) or deforming the laser shape from single- to two-color type. 
Our result indicates that THz-laser response can be a powerful tool to detect less-detectable natures of many-spin states in a broad class of magnets. 
%%%%%%%%%%%%%%%%%%%%%%%%%%%%%%%%%%%%
%  fig. 1  %%
%%%%%%%%

%%%%%%%%%%%%%%%%%%%%%%%%%%%%%%%%%%%%%%%%%%
%%%%%%%%%%%%%%%%%%%%%%%%%%%%%%%%%%%%%%%%%%
%%%%%%%%%%%%%%%%%%%%%%%%%%%%%%%%%%%%%%%%%%
%%%%%%%%%%%%%%%%%%%%
% Model and Method
%%%%%%%%%%%%%%%%%%%%
%\section{Model and Methods}
{\it Model and Methods}-- 
The target of this study is the Shastry-Sutherland model (SSM), whose Hamiltonian is %given by
\begin{align}
	\hat{\mathscr{H}}_\mathrm{SS}=
	J'\sum_\mathrm{\braket{i.j}}\hat{\bm{S}}_i\cdot\hat{\bm{S}}_j
	+J\sum_\mathrm{\braket{\braket{i.j}}}\hat{\bm{S}}_i\cdot\hat{\bm{S}}_j,
\label{eq:SS_SS}
\end{align}
where $\hat{\bm{S}}_j$ is the spin-$\frac{1}{2}$ operator on $j$-th site. 
As shown in Fig.~\ref{fig:model_SS}(a), $J$ and $J'$ are the intradimer and interdimer interactions, respectively. 
We set $J$ as a unit of energy and choose $J'/J=0.5$, where the ground state is the DSP that we mainly focus on. Under a DC magnetic field, the Zeeman term $\hat{\mathscr{H}}_\mathrm{Zeeman}=-\sum_i\bm{h}\cdot\hat{\bm{S}}_i$ is added. 
Here, $h_\alpha=g_\alpha\mu_\mathrm{B}H^\alpha,~(\alpha=x,~z)$, where $g_\alpha$ is the g-tensor, $\mu_\mathrm{B}$ is the Bohr magneton, and $H^\alpha$ is the magnetic field. 

\begin{figure}[t]
\begin{center}
\includegraphics[width=80mm]{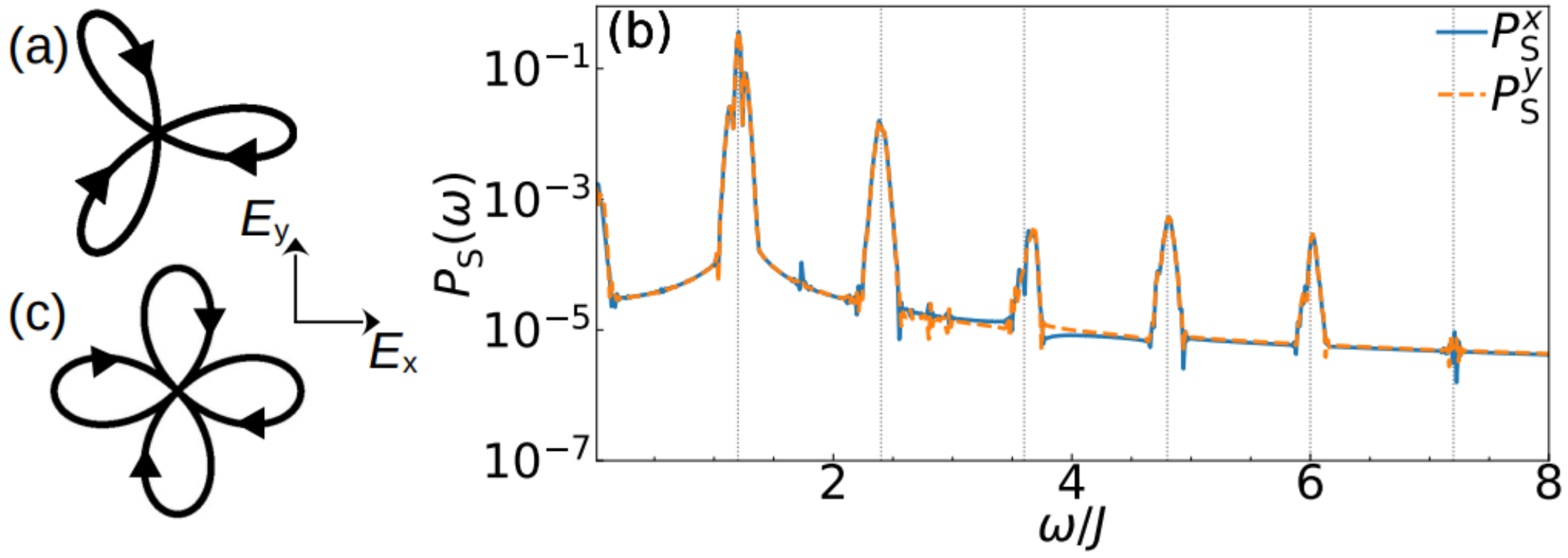}
\caption{
(a) Trajectory of the laser field of bicircular light [Eq.~(\ref{eq:bicircular}) in $l=2$] with 3-fold rotational symmetry. 
(b) %The electric polarization $\bm{P}_\mathrm{S}(\omega)$ under the bicircular light at $D=0$, $D'/J=0.015$ without a DC magnetic field.
Harmonic spectra $\bm{P}_\mathrm{S}(\omega)$ under the bicircular light in the SU(2) model with $D=0$, $D'=0$, and ${\bm H}={\bm 0}$.
Blue and orange lines respectively represent $P_\mathrm{S}^x(\omega)$ and $P_\mathrm{S}^y(\omega)$.
Vertical dotted lines represent integer multiples of $\omega_\mathrm{p}$. 
%Vertical dotted lines represent integer multiples of $\omega_\mathrm{p}$, with gray lines corresponding to multiples of 3 and black lines corresponding to the others. 
(c) Trajectory of the laser field of bicircular light [Eq.~(\ref{eq:bicircular}) in $l=3$] with 4-fold rotational symmetry. 
}
\label{fig:PSS_circularly_DSP_SS}
\end{center}
\end{figure}

In addition to the exchange interactions, the crystal structure of SCBO allows the emergence of Dzyaloshinskii-Moriya (DM) interactions~\cite{Dzyaloshinsky1958,Moriya1960}
%cite Dzyaloshinskii and Moriya papers
\begin{align}
	\hat{\mathscr{H}}_\mathrm{DM}=
	\sum_\mathrm{\braket{i.j}}\bm{D}'_{ij}\cdot(\hat{\bm{S}}_i\cross\hat{\bm{S}}_j)
	+\sum_\mathrm{\braket{\braket{i.j}}}\bm{D}_{ij}\cdot(\hat{\bm{S}}_i\cross\hat{\bm{S}}_j),
\label{eq:DM_SS}
\end{align}
where $\bm{D}'_{ij}$ and $\bm{D}_{ij}$ are the interdimer and intradimer DM vectors, respectively. 
Figure~\ref{fig:model_SS}(b) shows the ordering of i- and j-site spins in each DM term $\hat{\bm{S}}_i\cross\hat{\bm{S}}_j$; $\bm{D}'_{ij}=(0,0,D')$ for interdimer bonds, and $\bm{D}_{ij}=(0,D,0)$ $[(-D,0,0)]$ for A [B] intradimer bonds~\cite{Cepas2001,Miyahara2003,Miyahara2004,Miyahara2005,Kodama2005,Kakurai2005,Cheng2007,Mazurenko2008,Takigawa2010,Romhanyi2011,Chen2020}. 
For SCBO, the interdimer DM terms explain the splitting observed in the lowest branch of the triplon excitation in zero field~\cite{Cepas2001}, while the intradimer DM terms mix the ground singlet state with the triplon, which explains the anti-crossing observed in the experiment~\cite{Nojiri2003,Room2004}. 
When only $D'$ is finite, the system possesses U(1) symmetry, but the introducing $D$ destroys this symmetry. SCBO exhibits a structural transition at \SI{395}{K}~\cite{Sparta2001, Romhanyi2011}, 
below which the symmetry is lowered. 
The DM term with $D'$ ($D$) is relevant to the high and low  temperature phases (low temperature phase) of SCBO. 
We examine the U(1) breaking case, and the parameters are chosen to be $D'=0.015J$ and $D=0.025J$ 
%for the parameter choice 
except for Fig.~\ref{fig:PSS_circularly_DSP_SS} dealing with the SU(2) symmetric case. 

Next, we consider the light-matter coupling. The fundamental coupling in spin systems is the AC Zeeman interaction whose Hamiltonian is $\hat{\mathscr{H}}_\mathrm{Zeeman}(t)=-\sum_i\bm{h}_\mathrm{p}(t)\cdot\hat{\bm{S}}_i$, where $h^\alpha_\mathrm{p}(t)=g_\alpha \mu_{\rm B}H_{\rm p}^\alpha(t)$ and $\bm{H}_{\rm p}(t)$ is the laser magnetic field. In addition to it, multiferroic magnets including SCBO can be coupled to the electric field of the laser through magnetoelectric (ME) coupling~\cite{Kimura2003,Katsura2005,Mostovoy2006,Pimenov2006,Cheong2007,Takahashi2012,Tokura2014,Baierl2016,Arima2011,Miyahara2023}, 
which means the interaction between electric polarization $\hat{\bm{P}}$ and electron spin.  %degrees of freedom. 
The polarization $\hat{\bm{P}}$ interacts with the applied laser via the Hamiltonian 
$\hat{\mathscr{H}}_\mathrm{ME}(t)=-\bm{E}_\mathrm{p}(t)\cdot\hat{\bm{P}}$, where $\bm{E}_\mathrm{p}(t)$ is the electric field of laser. As a result, spins in multiferroics are coupled to the electric field of the laser as well. According to Ref.~\cite{Miyahara2023,Cepas2004,Room2004,Kimura2018}, SCBO has two types of ME couplings: The first is the exchange-striction type~\cite{Pimenov2006,Tokura2014} 
%cite https://www.nature.com/articles/nphys212
of ME coupling on the $J'$ bond, whose Hamiltonian is given by
\begin{align}
\hat{\bm{P}}_\mathrm{S}
=\sum_\mathrm{\braket{i.j}}\Pi_{\rm me}\bm{e}_{ij}\hat{\bm{S}}_i\cdot\hat{\bm{S}}_j,
\label{eq:MS}
\end{align}
where $\Pi_{\rm me}$ is the coupling constant and $\bm{e}_{ij}$ is the unit vector pointing from $i$ to $j$ sites on a $J'$ bond.  
The second ME interaction is the so-called inverse DM 
type~\cite{Katsura2005,Huvonen2009,Furukawa2010,Takahashi2012,Tokura2014} 
on the $J$ bond and the Hamiltonian~\cite{Miyahara2023} is  
\begin{align}
\hat{\bm{P}}_\mathrm{AS}
=\sum_\mathrm{\braket{\braket{i.j}}}D_{\rm me}\bm{e}_{ij}\cross\left(\hat{\bm{S}}_i\cross\hat{\bm{S}}_j\right),
\end{align}
where $D_{\rm me}$ is the coupling constant and $\bm{e}_{ij}$ is defined on a $J$ bond. 
For the laser fields $\bm{E}_\mathrm{p}(t)$ and $\bm{H}_\mathrm{p}(t)$, we adopt a Gaussian pulse ${\bm A}(t)={\bm A}\sin(\omega_\mathrm{p}(t-t_0))\exp(-(t-t_0)^2/2\sigma_\mathrm{p}^2)$, as shown in Fig.~\ref{fig:model_SS}(e), with the frequency $\omega_\mathrm{p}$ and the pulse width $\sigma_\mathrm{p}$ centered at time $t_0$. 
Here, $\bm A(t)=\bm{E}_\mathrm{p}(t)$ or $\bm{H}_\mathrm{p}(t)$, and $\bm A$ denote the amplitudes ${\bm E}_{\rm p}$ or ${\bm H}_{\rm p}$. 
%Unless otherwise noted, we irradiate linearly polarized light and calculate the electric polarization (the total magnetization) with components in the same direction as an electric (magnetic) field of the laser and then consider the dynamical symmetries in this case. 
Unless otherwise noted, when we consider the dynamics of the $\alpha$ component of electric polarization ${\hat P}^\alpha$ (magnetization $\hat{M}^\alpha=\sum_ig_\alpha\mu_\mathrm{B}\hat{S}_i^\alpha$), we apply an AC electric field $E_{\rm p}^\alpha(t)$ (magnetic field $H_{\rm p}^\alpha(t)$) with the same polarization direction to the SSM. 

The light-spin coupling is generally much smaller than the light-charge one, and it is difficult to detect the nonlinear optical responses in spin systems. 
However, as we already mentioned, we can apply sufficiently intense THz laser to materials with currently available techniques~\cite{Hirori2011,Sato2013,Hafez2016,Dhillon2017,Liu2017-2,Mukai2016,Kovalev2018,Evangelos2021,Alfred2023}. In multiferroics, 
the ME-coupling strengths of $\Pi_{\rm me}$ and $D_{\rm me}$ depend on the detail of materials and frequency, and 
they can be comparable to that of Zeeman interaction in materials with a strong enough ME coupling, i.e., $|\Pi_{\rm me}E_{\rm p}| \sim |D_{\rm me}E_{\rm p}|\sim |g_\alpha\mu_{\rm B}H_{\rm p}^\alpha|$. Hereafter, we suppose that the SSM has such a strong ME coupling. In fact, recent experiments show that SCBO has a large ME coupling of Eq.~(\ref{eq:MS})~\cite{Room2004,Giorgianni2023}. 
%{\color{red} Note that the state-of-the-art source, which separates purely the magnetic pulse from the light~\cite{Blanco2019,de2023} is devepoling.This can be powerful for the experiment in SCBO to demonstrate our prediction, especially seen in the total magnetization originating from the coupling between the magnetic field and spin.%, where the target is light-spin coupling rather than the dominant light-charge one.}

We note again that 
THz- and GHz-photons' energies are usually much smaller than the charge excitation gap of Mott insulators including SCBO, making the light-charge coupling irrelevant. 
In fact, THz-laser-driven magnetic modes without charge excitations have been detected in many experiments~\cite{Nicoletti2016,Kampfrath2011,Kampfrath2013,Vicario2013,Baierl2016-2,Mukai2016,Lu2017-2,Bossini2019,Li2020,Walowski2016,Salen2019,Safin2020,Pimenov2006,Takahashi2012,Kubackaand2014,Wang2018,Zhang2023}.
Therefore, it is sufficient to consider only light-spin couplings without light-charge ones when we apply THz or GHz waves to the SSM.

We calculate expectation values of the electric polarization $\hat{\bm P}$ and the magnetization $\hat{\bm M}$
%$\hat{M}^\alpha=\sum_ig_\alpha\mu_\mathrm{B}\hat{S}_i^\alpha$ 
as a function of time, using %ED 
the numerical Lanczos method~\cite{Park1986,Mohankumar2006,Wang2017-2} in the $16$-site cluster,  which is invariant under the operations for $\pi/2$ rotation  
 of lattice~\cite{Furukawa2011}.
The simulation is done in a finite time range $[0,t_\mathrm{max}]$, and the dissipation effect is not included in the ED method. 
Therefore, we should introduce a 
window function $w(t)=0.54-0.46\cos(2\pi t/(t_\mathrm{max}-1))$ to obtain realistic harmonics spectra via an effective Fourier transformation~\cite{Prabhu2014,Prajoy2014}. 
Namely, for the expectation value of a physical quantity $O(t)$, we define its spectrum as $O(\omega)=\int_0^{t_{\rm max}}O(t)w(t)e^{i\omega t}\dd t$. We will use symbols $P^\alpha_{\rm S}(\omega)$, $P^\alpha_{\rm AS}(\omega)$ and $M^\alpha(\omega)$ as the effective Fourier transforms of $\langle {\hat P}^\alpha_{\rm S}(t)\rangle$, $\langle {\hat P}^\alpha_{\rm AS}(t)\rangle$, and $\langle {\hat M}^\alpha(t)\rangle$, respectively. 
In the simulations, 
the strengths of light-matter couplings are fixed to
$|\Pi_{\rm em}E_{\rm p}|=0.1J$, $|D_{\rm em}E_{\rm p}|=0.1J$ or $|g_\alpha\mu_{\rm B}H_{\rm p}^\alpha|=0.2J$ unless otherwise noted. One can approach these large values by using a THz pulse with a strong electric field of $\sim 1$MV/cm.

%%%%%%%%%%%%%%%%%%%%%%%%%%%%%%%%%%%%%%%%%%%
%%%%%%%%%%%%%%%%%%%%%%%%%%%%%%%%%%%%%%%%%%%
%%%%%%%%%%%%%%%%%%%%%%%%%%%%%%%%%%%%%%%%%%%
%%%%%%%%%%%%%%%%%%%%
% Results
%%%%%%%%%%%%%%%%%%%%
%\section{Results}
{\it Results}-- 
Below we show %some 
several important findings of our numerical computation, features of THz-pulse driven dynamics in the SSM. More details are in Supplemental Material (SM)~\cite{SM}. 

\begin{figure}[t]
\begin{center}
\includegraphics[width=\columnwidth]{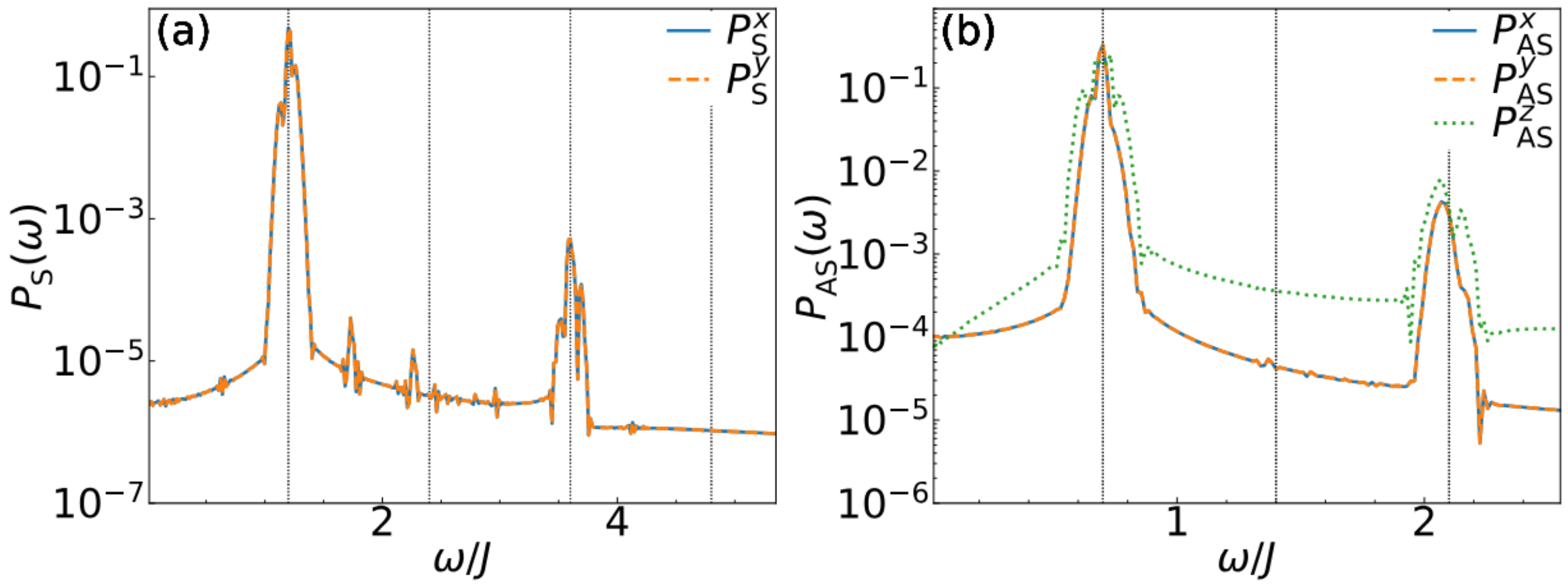}
\caption{
Computed HG spectra of (a) $\bm{P}_\mathrm{S}(\omega)$ and (b) $\bm{P}_\mathrm{AS}(\omega)$ under zero field ${\bm H}={\bm 0}$. 
Blue and orange lines in both figures represent the $x$ and $y$ components of the electric polarizations, respectively, while the green line in (b) shows the $z$ component of $\bm{P}_\mathrm{AS}(\omega)$.
Vertical dotted lines represent integer multiples of $\omega_\mathrm{p}$. 
For $J=1$ meV, $\omega=J$ corresponds to $2.42\times 10^{11}$ Hz.
%Vertical dotted lines represent integer multiples of $\omega_\mathrm{p}$, with black (gray) lines corresponding to odd- (even-) harmonics. 
}
\label{fig:P_h0_DSP_SS}
\end{center}
\end{figure}

%%%%%%%%%%%%%%%%%%%%%%%%%%%%%%%%%%%%%%%%%%%
%%%%%%%%%%%%%%%%%%%%%%%%%%%%%%%%%%%%%%%%%%%
%\subsection{The resonance peaks for electric polarizations}
{\it A: Electric polarizations}-- 
In the main text, we concentrate on the dynamics driven by ME couplings. That driven by AC Zeeman coupling is in the SM~\cite{SM}.
The main mode excited by $\hat{\bm{P}}_\mathrm{S}$ is the triplon bound state with $S=0$ (resonance frequency $\omega/J\sim 1.2$), while that by $\hat{\bm{P}}_\mathrm{AS}$ is the single triplon (resonance frequency $\omega/J\sim 0.7$)~\cite{Totsuka2001,Miyahara2023}. 
Figure~\ref{fig:P_h0_DSP_SS} shows the harmonic spectra $P^\alpha_\mathrm{S}(\omega)$ ($\alpha=x,~y$) driven by a light-matter coupling $\hat{\mathscr{H}}_\mathrm{ME}(t)=-E^\alpha_\mathrm{p}(t){\hat P}^\alpha_\mathrm{S}$ and $P^\alpha_\mathrm{AS}(\omega)$ ($\alpha=x,~y,~z$) by $-E^\alpha_\mathrm{p}(t){\hat P}^\alpha_\mathrm{AS}$ in the model with  $\bm H=\bm 0$. 
The laser frequency $\omega_\mathrm{p}$ is chosen to their resonance frequencies. 
%In addition to 
Besides the fundamental harmonic, third-order one is also observed, but even-order responses do not appear. 
The fundamental (linear) responses of $P^\alpha_\mathrm{S}(\omega_{\rm p})$ and $P^\alpha_\mathrm{AS}(\omega_{\rm p})$ are consistent with those of a numerical study for a larger system~\cite{Miyahara2023}, which means that our results for 16-site clusters can capture the thermodynamic nature in a semi-quantitative level. Figure~\ref{fig:P_h0_DSP_SS} also indicates that nonlinear third harmonics can be observed by utilizing 
currently available intense THz laser. 
%besides the linear response. 
When a static magnetic field is applied, we verify that the spectral peak of $\hat{\bm{P}}_\mathrm{AS}$ exhibits the Zeeman splitting, while that of $\hat{\bm{P}}_\mathrm{S}$ does not. 
This is a clear evidence for detecting triplons and triplon pairs in Fig.~\ref{fig:P_h0_DSP_SS}. The splitting is explained in more detail in Supplemental Material (SM)~\cite{SM}. 
Here we remark two things: The first is that tirplons and triplon pairs with only a certain wave vector (typically ${\bm k}={\bm 0}$) are detected by the HG because applied waves have a long wavelength and may be viewed as a spatially uniform field for electron spins. Secondly, small spikes in the HG spectra are expected to be negligible in the real materials (for more detail, see SM).

%%%%%%%%%%%%%%%%%%%%%%%%%%%%%%%%%%%%%%%%%%%
%  fig. 2  %%
%%%%%%%%

\begin{figure}[t]
\begin{center}
\includegraphics[width=\columnwidth]{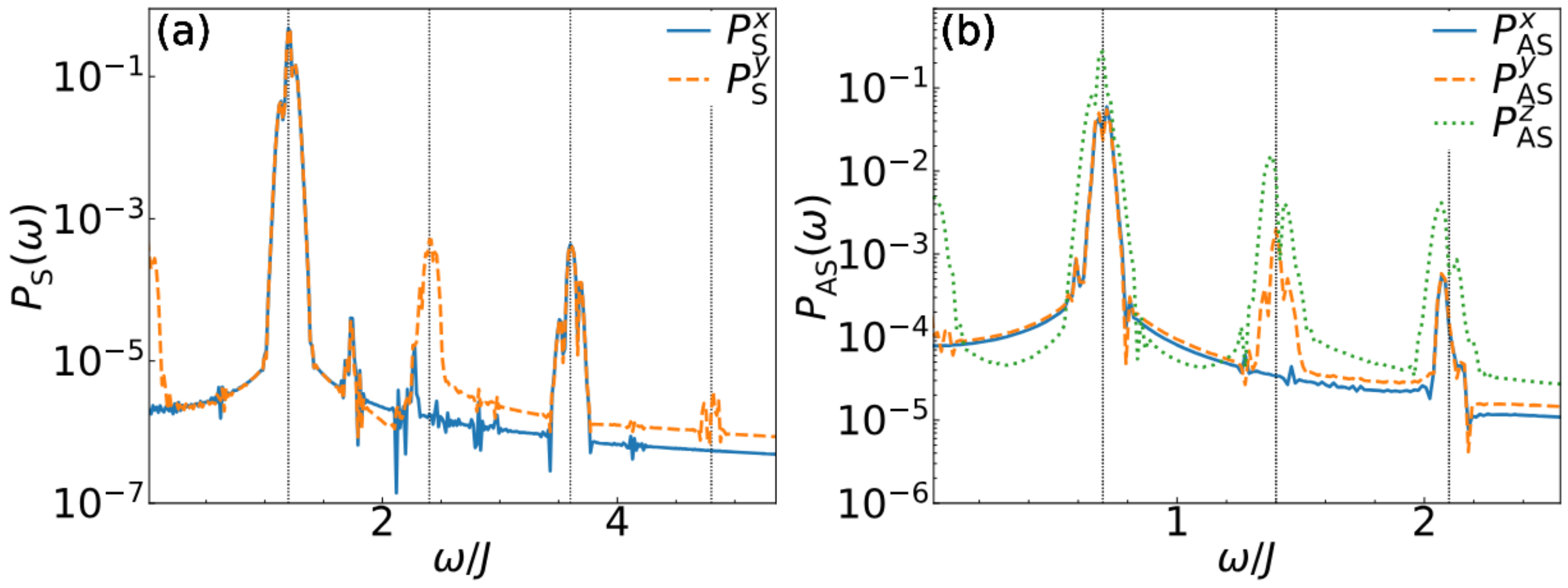}
\caption{
HG spectra of (a) $\bm{P}_\mathrm{S}(\omega)$ and (b) $\bm{P}_\mathrm{AS}(\omega)$ calculated with $h_x/J=0.1$. 
Blue and orange lines in all figures represent the $x$ and $y$ components of $\bm{P}_\mathrm{S}(\omega)$ or $\bm{P}_\mathrm{AS}(\omega)$, respectively, while the green line in (b) shows the $z$ component of $\bm{P}_\mathrm{AS}(\omega)$.
Vertical dotted lines represent integer multiples of $\omega_\mathrm{p}$. 
%The meaning of the vertical dotted line is the same as in Fig.~\ref{fig:P_h0_DSP_SS}. 
}
\label{fig:P_DSP_SS}
\end{center}
\end{figure}

%%%%%%%%%%%%%%%%%%%%%%%%%%%%%%%%%%%
%\section{Dynamical symmetries in the DSP}
{\it B: Dynamical symmetries in DSP}--
Next, we consider the selection rule of the $n$-th order harmonics of the SSM. To this end, it is %very 
important to find a dynamical symmetry of the periodically driven system, in which the laser pulse should be replaced with an ideal continuous wave ($\sigma_{\rm p}\to\infty$). 
If a time-periodic system with a Hamiltonian $\hat{\mathscr{H}}(t)=\hat{\mathscr{H}}(t+T_\mathrm{p})$ ($T_\mathrm{p}$ is the time period) satisfies the relation $\hat{U}\hat{\mathscr{H}}(t+T_\mathrm{p}/m)\hat{U}^\dag=\hat{\mathscr{H}}(t)$ with $m\in\{2,3,\cdots\}$ and $\hat{U}$ being a symmetry operator, one says that the system possesses the dynamical symmetry associated with the symmetry of $\hat{U}$. 
In laser-driven systems, $T_\mathrm{p}=2\pi/\omega_\mathrm{p}$ stands for the period of the laser.
%and $\omega_\mathrm{p}$ is the laser frequency. 
One can often find a selection rule of harmonic spectra in dynamically symmetric systems. For more details of the dynamical symmetry, see SM~\cite{SM}.
We can find dynamical symmetries of $\hat{U}\hat{\mathscr{H}}(t)\hat{U}^{-1}=\hat{\mathscr{H}}(t+T_\mathrm{p}/2)$ when the total Hamiltonian is given by 
$\hat{\mathscr{H}}(t)=\hat{\mathscr{H}}_\mathrm{SS}+\hat{\mathscr{H}}_\mathrm{DM}+\hat{\mathscr{H}}_\mathrm{ME}(t)$, and the symmetry operator $\hat{U}$ is chosen to be $C_2(z)$ or $\sigma_{yz}$ for $E_{\rm p}^x$, $C_2(z)$ or $\sigma_{zx}$ for $E_{\rm p}^y$, and $S_4$ for $E_{\rm p}^z$~\cite{Jacobs2005,Dresselhaus2007,Romhanyi2011}. 
However, except for $\sigma_{yz}$ ($\sigma_{zx}$), they are broken when a DC magnetic field $h_{x(y)}$ (together with an intradimer DM interaction) is introduced. 
The spectra ${\bm P}_\mathrm{S(AS)}(\omega)$ with the DC magnetic field $h_x$ are shown in Fig.~\ref{fig:P_DSP_SS}. 
Figure~\ref{fig:P_h0_DSP_SS} shows that in the cases of $h_x=0$, there is no dynamical symmetry breaking and any even-order peak disappears, whereas Fig.~\ref{fig:P_DSP_SS} shows that under a magnetic field $h_x\neq 0$, the dynamical symmetry about $P_\mathrm{S}^y(\omega)$ and $P_\mathrm{AS}^{y,z}(\omega)$ are broken (that of $P_\mathrm{S(AS)}^x(\omega)$ survives) and the even-order responses for these polarizations are observed.  
The selection rule for even-order harmonics in electron systems is always governed by an inversion symmetry, while we emphasize that in the case of magnetic-excitation induced harmonics~\cite{Ikeda2019,Ikeda2020,Kanega2021,Kanega2024,Zhang2023}, 
several selection rules associated with different symmetries can appear depending on the spin-light couplings and static symmetries of the systems. The above selection rule also indicates that an external magnetic field and its direction can control the breaking of dynamical symmetries. More details of selection rules are summarized in SM.  
%%%%%%%%%%%%%%%%%%%%%%%%%%%%%%%%%%%%%%%%%%%%
%  fig. 4  %%
%%%%%%%%

%%%%%%%%%%%%%%%%%%%%%%%%%%%%%%%%%%%%%%%%%%%%%%%
%%%%%%%%%%%%%%%%%%%%%%%%%%%%%%%%%%%%%%%%%%%%%%%
%%%%%%%%%%%%%%%%%%%%%%%%%%%%%%%%%%%%%%%%%
%%%%%%%%%%%%%%%%%%%%%%%%%%%%%%%%%%%%%%%%%
{\it C: Two-color laser}--
In the above paragraphs, we show that the harmonic spectra can be controlled by a static term in the Hamiltonian like a DC magnetic field and a DM interaction. However, 
dynamical symmetries can also be dependent on the shape of the laser field. 
Here, we consider the possibility of controlling the %harmonic 
spectra by changing the laser wave shape. 
%In the case of a single laser, this symmetry is determined by the polarization of light, but variations of the symmetry increase when multiple lasers are combined. 
%Next, we show a result under circularly polarized light irradiation in order to reduce any symmetry using lasers instead of the DC magnetic field or antisymmetric interactions. 
For this purpose, we deal with a bicircular laser, a combination of a left circularly polarized light (CPL) with frequency $\omega_{\rm p}$ and a right CPL with $\ell\omega_{\rm p}$. 
The electric field $\bm{E}_\mathrm{p}(t)$ of the bicircular laser is given by 
\begin{align}
\label{eq:bicircular}
	&\bm{E}_\mathrm{p}(t)
	=\bm{E}_\mathrm{p}e^{-(t-t_0)^2/2\sigma_\mathrm{p}^2}	\left(\bm{E}_\mathrm{p1}(t)+\bm{E}_\mathrm{p2}(t)\right)/\sqrt{2}, 
	\notag\\
	&\bm{E}_\mathrm{p1}(t)
	=\left[\cos(\omega_\mathrm{p}(t-t_0))\hat{\bm{e}}^x+\sin(\omega_\mathrm{p}(t-t_0))\hat{\bm{e}}^y\right], 
	\notag\\
	&\bm{E}_\mathrm{p2}(t)
	=\left[\cos(\ell\omega_\mathrm{p}(t-t_0))\hat{\bm{e}}^x-\sin(\ell\omega_\mathrm{p}(t-t_0))\hat{\bm{e}}^y\right], 
\end{align}
where the $\bm{E}_\mathrm{p1}(t)$ [$\bm{E}_\mathrm{p2}(t)$] is the field of the left [right] CPL, and $\hat{\bm{e}}^x$ ($\hat{\bm{e}}^y$) is the unit vector of $x$ ($y$) direction. 
The integer $\ell$ represents the multiple of the laser frequency with respect to the first laser.
Figure~\ref{fig:PSS_circularly_DSP_SS} (a) depicts the trajectory of the bicircular laser with $\ell=2$ and it leads to a different spacial symmetry from linearly or circularly polarized light. 
These sorts of multiple-color lasers are applicable with the current laser techniques~\cite{Fleischer2014,Kfir2015,Eichmann1995,Baykusheva2016,Ogawa2023}. 
Figure~\ref{fig:PSS_circularly_DSP_SS} (b) shows the spectra $P^{x,y}_\mathrm{S}(\omega)$ under this bicircular light with $\ell=2$ in the SU(2)-symmetric SSM with $D=0$, $D'=0$ and ${\bm H}={\bm 0}$. 
The SSM does not have the three-fold rotational symmetry, in other words, incompatible with the symmetry of the bicircular light. 
It means that no symmetry operator $\hat{U}$ satisfies the dynamical symmetry $\hat{U}\hat{\mathscr{H}}(t)\hat{U}^{-1}=\hat{\mathscr{H}}(t+T_\mathrm{p}/\ell)$ for $\ell=2$ even though the total Hamiltonian keeps the SU(2) symmetry without DM interactions. 
As a result, all the integer order responses, including the peaks with multiples of 3, appear as shown in 
Fig.~\ref{fig:PSS_circularly_DSP_SS}(b). 
Choosing the two lasers to have a four-fold rotational symmetry shown in Figure~\ref{fig:PSS_circularly_DSP_SS} (c), dynamical symmetry is not broken, and even-order does not appear.
We deal with its detail in 
%Section C %~\ref{sec:BCL4R_SS} of 
the SM~\cite{SM}. 
%%%%%%%%%%%%%%%%%%%%%%%%%%%%%%%%%%%%%%%%%%%%%%%%
%  fig. 5  %%
%%%%%%%%

\begin{figure}[t]
\begin{center}
\includegraphics[width=\columnwidth]{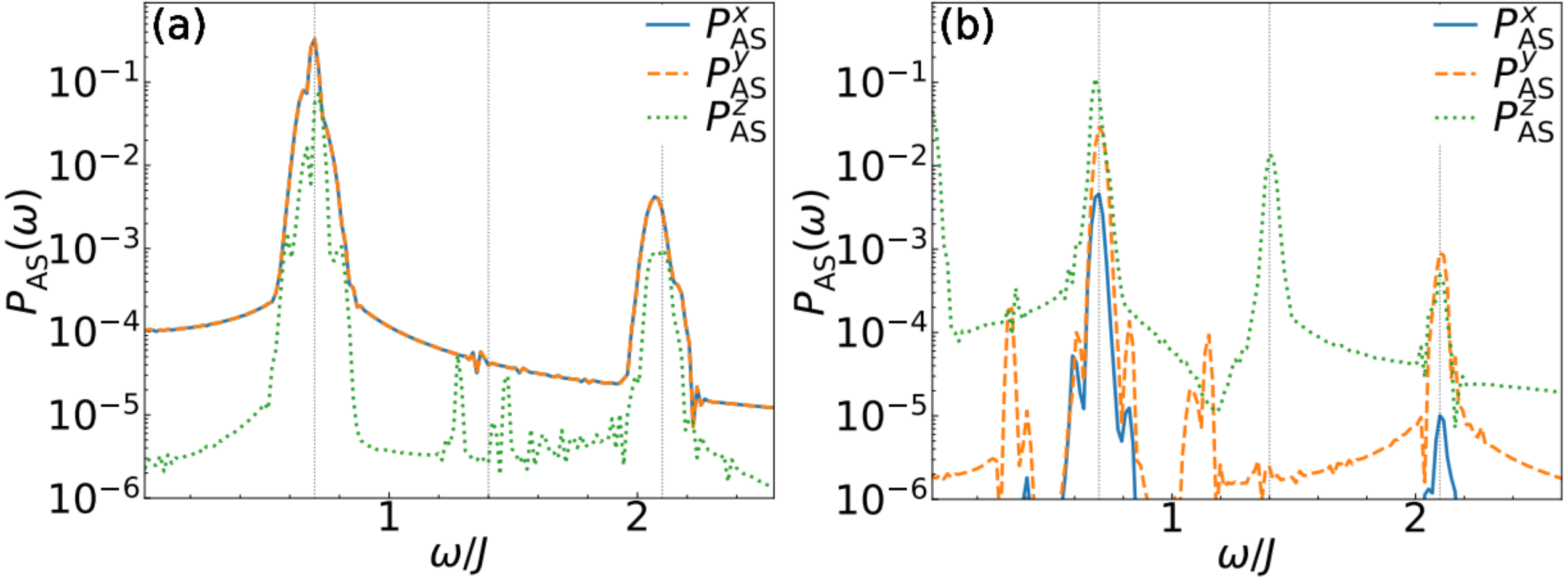}
\caption{
$\bm{P}_\mathrm{AS}(\omega)$ (a) in the DSP with $h_z/J=0.1$ and $h_\mathrm{s}=0$  and (b) in the $\frac{1}2$ PP with $h_z/J=1.5$ and $h_\mathrm{s}/J=0.01$.
Even-order harmonic peaks of $P_\mathrm{AS}^z$, which does not exist in the DSP, appear in the PP. 
Blue, orange, and green lines in both figures represent the $x$, $y$, and $z$ components of $\bm{P}_\mathrm{AS}(\omega)$.
Vertical dotted lines represent integer multiples of $\omega_\mathrm{p}$. 
%The meaning of the vertical dotted line is the same as in Fig.~\ref{fig:P_h0_DSP_SS}. 
}
\label{fig:PAS_12PP_SS}
\end{center}
\end{figure}

%%%%%%%%%%%%%%%%%%%%%%%%%%%%%%%%%%%%%%%%%%%%%%%
%%%%%%%%%%%%%%%%%%%%%%%%%%%%%%%%%%%%%%%%%%%%%%%
%%%%%%%%%%%%%%%%%%%%%%%%%%%%%%%%%%%
%\section{Comparison between the DSP and other phases about dynamical symmetries}\label{subsec:other_SS}
%%%%%%%%%%%%%%%%%%%%%%%%%%%%%%%%%%%%%
%%%%%%%%%%%%%%%%%%%%%%%%%%%%%%%%%%%%%
%{\it D: Plaquette and half-plateau phases}-- 
{\it D: Ordered phases}-- 
So far, we have focused on the THz-laser-driven dynamics in the DSP, which is relevant to SCBO. 
As we mentioned in the Introduction, 
if we introduce a strong enough perturbation like a pressure or a magnetic field, SCBO enters other phases accompanying a spontaneous symmetry breaking. Here, we analyze the harmonic spectra 
%in two ordered phases, the empty plaquette phase (EPP) and 1/2 plateau phase (PP) [see Fig.~\ref{fig:model_SS} (c) and (d)]. 
in the $\frac{1}{2}$ plateau phase (PP) [see Fig.~\ref{fig:model_SS} (d)]. 
In the PP, two sorts of dimers respectively have different values of magnetizations, and as a result, the $S_4$ symmetry is spontaneously broken~\cite{SM}. 
To realize this phase %with strongly magnetized and less magnetized dimers 
in our finite 16-site cluster, we apply a small alternating magnetic field to bonds A and B as the mean field~\cite{Czarnik2021}: Namely, we add a perturbation $\hat{\mathscr{H}}_\mathrm{bias}=-h_\mathrm{s}\left(\sum_{i\in\mathrm{A}}\hat{S}_i^z-\sum_{i\in\mathrm{B}}\hat{S}_i^z\right)$, which is equivalent to the order parameter in this phase.
Recalling that the dynamical symmetry with $\hat{U}=S_4$ is relevant to $E_{\rm p}^z$, it is expected that $P^z_\mathrm{AS}(\omega)$ shows even-order responses in the PP with light-matter coupling $\hat{\mathscr{H}}_\mathrm{ME}(t)=-E_{\rm p}^z(t)P^z_\mathrm{AS}(\omega)$.
The estimated electric polarizations $P^\alpha_\mathrm{AS}(\omega)$ %($\alpha=x,~y,~z$) 
in the DSP and PP are depicted in Fig.~\ref{fig:PAS_12PP_SS} (a) and (b), respectively. 
We numerically verify that for $J'/J=0.5$, the $\frac{1}2$ PP exists in the range $1.1\alt h_z/J\alt 1.9$, and therefore we set $h_z/J=1.5$ to estimate the harmonic spectrum in the PP.
%$D/J=0.025$, $D'/J=0.015$ regarding the DM term, and $h_z/J=0.1$ for the DC magnetic field along the $z$ axis. 
Figure~\ref{fig:PAS_12PP_SS} shows that in the PP, even-order harmonics of $P_\mathrm{AS}^z(\omega)$ appear, whereas there is no even-order peak in the DPS. This result indicates that the selection rule for harmonic spectra changes by ordering patterns. It means that we can detect important information about magnetic order and symmetry breakdowns by observing THz-laser-driven harmonic spectra. 
Similar breakings of dynamical symmetries can occur in other ordered phases such as plaquette ordered phases [Fig.~\ref{fig:model_SS} (c)] and spin-nematic phase predicted in Ref.~\cite{Batista2018}. These dynamical symmetry breakdowns and the change of the selection rules due to the emergence of orders will be explained in SM~\cite{SM}. 
%Note that many high-energy modes fall off in the PP at Zeeman splitting, and numerous fine peaks are seen in $P^x_\mathrm{AS}$ and $P^y_\mathrm{AS}$.
%The situation is the same for $P^z_\mathrm{AS}$, but the overall spectral intensity is large over all frequencies, which is thought to drown out these fine peaks.

%%%%%%%%%%%%%%%%%%%%%%%%%%%%%%%%%%%%%%%%%%%%%%%%%%%
%  fig. 6  %%
%%%%%%%%

%%%%%%%%%%%%%%%%%%%%%%%%%%%%%%%%%%%%%%%%%%%%%%%
%%%%%%%%%%%%%%%%%%%%%%%%%%%%%%%%%%%%%%%%%%%%%%%
%%%%%%%%%%%%%%%%%%%%%%%%%%%%%%%%%%%%%%%%%%%%%%%
%%%%%%%%%%%%%%%%%%%%
% Summary
%%%%%%%%%%%%%%%%%%%%
%\section{Summary}
{\it Summary}--
In summary, we have numerically investigated the THz-laser-driven nonlinear responses of the SSM with/without the DM interactions, which could be observed in SCBO. 
Triplons are detectable in the electric polarization $\bm{P}_\mathrm{AS}(\omega)$, while triplon pairs 
%with $S=0$ 
are in $\bm{P}_\mathrm{S}(\omega)$. 
We demonstrate that the selection rule of the HG spectra in the SSM can be controlled by manipulating a DC magnetic field, DM interactions, and the shape of laser. Moreover, we find that the rule is also changed depending on the ordering patterns in the SSM. 

In magnets, the SU(2) spin-rotational and spatial symmetries are often broken in various manners, depending on magnetic orders and anisotropies. 
Our results indicate that THz or GHz wave driven HG spectra are powerful in obtaining the important properties of quasiparticles, magnetic orderings, and the sorts of light-matter couplings in broad phases of magnets. This is in contrast with semiconductors and metals, in which the HG spectra are usually used to detect only the inversion symmetry breaking.

%%%%%%%%%%%%%%%%%%%%
% Acknowledgements
%%%%%%%%%%%%%%%%%%%%

%\section*{Acknowledgments}
\begin{acknowledgments}
The authors would like to thank Shunsuke Furukawa for his valuable comments and support for the numerical method. 
This work was supported by KAKENHI (Grant No. JP20H01849), a Grant-in-Aid for Scientific Research on Innovative Areas “Evolution of Chiral Materials Science using Helical Light Fields” (Grants No. JP22H05131 and No. JP23H04576) from JSPS of Japan, and  JST, CREST Grant Number JPMJCR24R5, Japan. 
M.U. acknowledges the support by JST SPRING (Grant No. JPMJSP2109).
The numerical diagonalization program is based on the package TITPACK ver.2. 
\end{acknowledgments}

%%%%%%%%%%%%%%%%%%%%
% References
%%%%%%%%%%%%%%%%%%%%

\bibliography{draft.bbl}
%\bibliography{References}

\end{document}

%%%%%%%%%%%%%%%%%%%%
% Title and authors
%%%%%%%%%%%%%%%%%%%%

\title{Supplemental Material: Triplons, triplon pairs and dynamical symmetries in laser-driven Shastry-Sutherland magnets}
\author{Mina Udono}
\author{Masahiro Sato}
\affiliation{Department of Physics, Chiba University, Chiba 263-8522, Japan}
\date{\today}

\maketitle

In the Supplemental Material (SM), 
we will discuss some details of THz-laser-driven harmonic spectra in the Shastry-Sutherland model (SSM). 
As in the main text, we mainly focus on the U(1) breaking case with both interdimer and intradimer DM interactions. We fix the values of the DM coupling constants $D'/J=0.015$ and $D/J=0.025$ as in the main text.

%%%%%%%%%%%%%%%%%%%%%%%%%%%%%%%%%%%%
%%%%%%%%%%%%%%%%%%%%%%%%%%%%%%%%%%%%
%%%%%%%%%%%%%%%%%%%%%%%%%%%%%%%%%%%%
\subsection{Symmetry operations and selection rules for harmonic spectra}
In the first section of the SM, we briefly review the dynamical symmetry in laser-driven systems~\cite{Alon1998,Neufeld2019,Ikeda2019,Ikeda2020,Kanega2021,Kanega2024}, and then we explain the relationship between the dynamical symmetry and the corresponding selection rule for the spectra of high harmonic generation ~\cite{Ikeda2019,Ikeda2020,Kanega2021,Kanega2024}. As we mentioned in the main text, the usual laser electric field for harmonic generation (HG) is a pulse type:
\begin{align}
    {\bm E}_{\rm p}(t) ={\bm E}_{\rm p}\sin(\omega_{\rm p}(t-t_0))\exp(-(t-t_0)^2/\sigma_{\rm p}^2),
\end{align}
where ${\bm E}_{\rm p}$ is the amplitude of the electric field, 
$\omega_{\rm p}$ is the laser frequency, and $\sigma_{\rm p}$ denotes the length of laser pulse. 
The selection rule exactly holds when the applied laser is a continuous wave ($\sigma_{\rm p}\to \infty$). 
In this section, we consider this ideal setup of $\sigma_{\rm p}\to \infty$. The resulting selection rule can be applied even in real setups with a finite $\sigma_{\rm p}$ in a practical sense. 

In the following, as a specific example, we address the $\alpha$ component ($\alpha=x$, $y$ or $z$) of an electric polarization $\hat{P}^\alpha$ in a system under an oscillating electric field $E_\mathrm{p}^\alpha(t)$. However, the same argument applies to the magnetization dynamics in a system under an oscillating magnetic field $H_\mathrm{p}^\alpha(t)$. 
%The symbol $\alpha$ means a direction $x,~y,~z$, 
The index $\alpha$ will be omitted for simplicity in the rest of this section. Hereafter, $\hat{P}^\alpha$ is abbreviated as $\hat{P}$. 

First, we define a standard static symmetry for a unitary operator $\hat{U}$ (see Table~\ref{tb:symmetry_SS}) by the relation 
\begin{align}
\hat{U}\hat{\mathscr{H}}\hat{U}^\dag
=\hat{\mathscr{H}}
\label{eq:symmetry_SS}
\end{align}
where the Hamiltonian of the SSM is given by $\hat{\mathscr{H}}=\hat{\mathscr{H}}_\mathrm{SS}+\hat{\mathscr{H}}_\mathrm{DM}$ [Eqs. (1) %\eqref{eq:SS_SS}
and (2) %\eqref{eq:DM_SS}
in the main text]. 
If $\hat{S}^\alpha$ is invariant under the symmetry operation $\hat{U}$, 
%which is written in red text (yellow background box) in Table~\ref{tb:symmetry_SS}, 
Eq.~\eqref{eq:symmetry_SS} still holds even after adding a Zeeman term $\hat{\mathscr{H}}_\mathrm{Zeeman}=-\sum_ih_\alpha\hat{S}_i^\alpha$ to $\hat{\mathscr{H}}$. 

First, we consider a linearly polarized laser, whose electric field obeys $E_\mathrm{p}(t+T_\mathrm{p}/2)=-E_\mathrm{p}(t)$ with $T_\mathrm{p}=2\pi/\omega_\mathrm{p}$ being the period of the laser. 
From this, when the polarization $\hat{P}$ (see a circle in Table~\ref{tb:symmetry_SS}) 
satisfies the relation 
\begin{align}
	\hat{U}\hat{P}\hat{U}^\dag=-\hat{P}, 
\label{eq:quantity_SS}
\end{align}
then a dynamical symmetry 
\begin{align}
	\hat{U}\hat{\mathscr{H}}(t+T_\mathrm{p}/2)\hat{U}^\dag=\hat{\mathscr{H}}(t), 
\label{eq:dynamical_SS}
\end{align}
holds in the laser-driven system of the time-dependent Hamiltonian $\hat{\mathscr{H}}(t)=\hat{\mathscr{H}}+\hat{\mathscr{H}}_\mathrm{ME}(t)$, where $\hat{\mathscr{H}}_\mathrm{ME}(t)$ is the laser-induced ME coupling. 
Substituting Eq.~\eqref{eq:dynamical_SS} into the Sch\"odinger equation $i(\partial/\partial t)\ket{\psi(t)}=\hat{\mathscr{H}}(t)\ket{\psi(t)}$, and acting $\hat{U}^\dag$ from the left, we get $i\frac{\partial}{\partial t}\hat{U}^\dag\ket{\psi(t)}=\hat{\mathscr{H}}(t+T_\mathrm{p}/2)\hat{U}^\dag\ket{\psi(t)}$. 
Through a time shift $t\rightarrow t+T_\mathrm{p}/2$, we find that $\hat{U}^\dag\ket{\psi(t+T_\mathrm{p}/2)}$ also follow the Sch\"odinger equation for the Hamiltonian $\hat{\mathscr{H}}(t)$ because of the periodicity $\hat{\mathscr{H}}(t+T_\mathrm{p})=\hat{\mathscr{H}}(t)$. 
Therefore, we have 
$\hat{U}^\dag\ket{\psi(t+T_\mathrm{p}/2)}\propto\ket{\psi(t)}$ if there is no degeneracy. Combining it and Eq.~\eqref{eq:quantity_SS}, we obtain $P(t)=-P(t+T_\mathrm{p}/2)$, where $P(t)=\langle \hat P(t)\rangle$. This equation directly leads to the absence of even-order harmonics. It is proved as follows. The $n$-th harmonic spectrum $P(n\omega_\mathrm{p})$ is computed as 
\begin{align}
	P(n\omega_\mathrm{p})
	&=\frac{1}{T_\mathrm{p}}\int_0^{T_\mathrm{p}}P(t)e^{in\omega_\mathrm{p}t}\dd t
	\notag\\
	&=\frac{1}{T_\mathrm{p}}\int_{-\frac{T_\mathrm{p}}{2}}^{\frac{T_\mathrm{p}}{2}}
	P\left(t+\frac{T_\mathrm{p}}{2}\right)e^{in\omega_\mathrm{p}\left(t+\frac{T_\mathrm{p}}{2}\right)}\dd t
	\notag\\
	&=-e^{in\pi}\frac{1}{T_\mathrm{p}}\int_{-\frac{T_\mathrm{p}}{2}}^{\frac{T_\mathrm{p}}{2}}
	P(t)e^{in\omega_\mathrm{p}t}\dd t
	\notag\\
	&=-e^{in\pi}P(n\omega_\mathrm{p}). 
 \label{eq:selection}
\end{align}
Thus, we arrive at $P(2m\omega_\mathrm{p})=0~(m\in\mathbb{Z})$.

In the case of the bicircular laser with 3-fold rotational symmetry [see Fig.4 (a) of the main text], the oscillating electric field $\bm{E}_\mathrm{p}(t)$ satisfies $\bm{E}_\mathrm{p}(t+T_\mathrm{p}/3)=R(2\pi/3)\bm{E}_\mathrm{p}(t)$, where 
\begin{align}
    R(\theta)=
\left(\begin{array}{cc}
\cos\theta & -\sin\theta \\
\sin\theta & \cos\theta \\
\end{array}\right)
\end{align}
is the two-dimensional rotation matrix by $\theta$. 
Therefore, if there exists a symmetry operation satisfying $\hat{U}\hat{\bm{P}}\hat{U}^\dag=R(2\pi/3)\hat{\bm{P}}$ for the polarization vector $\hat{\bm{P}}=\hat{P}^x\bm{e}^x+\hat{P}^y\bm{e}^y$, we obtain the relation $\bm{P}(3m\omega_\mathrm{p})=0~(m\in\mathbb{Z})$ through a similar argument to Eq.~(\ref{eq:selection}). It means that $3m$-th order harmonics all disappear. 
In the present SSM, however, there is no symmetry operation $\hat{U}$ that fulfills the equality $\hat{U}\hat{\bm{P}}\hat{U}^\dag=R(2\pi/3)\hat{\bm{P}}$, so that the appearance of $3m$-th order harmonics is generally allowed [see Fig.~\ref{fig:PSS_circularly_DSP_SS} (b) in the following section]. 

On the other hand, if we consider the SSM irradiated by another bicircular laser with $\bm{E}_\mathrm{p}(t+T_\mathrm{p}/4)=R(\pi/2)\bm{E}_\mathrm{p}(t)$ [see Fig.~\ref{fig:PSS_circularly_DSP_SS} (a) in the following section], 
we can find the dynamical symmetry of $\hat{U}$ satisfying  $\hat{U}\hat{\bm{P}}\hat{U}^\dag=R(\pi)\hat{\bm{P}}$ because the SSM has the symmetry $C_2(z)$ (i.e., a 2-fold rotational symmetry). 
Therefore, for a bicircular laser with 4-fold rotational symmetry, we obtain $\bm{P}(2m\omega_\mathrm{p})=0~(m\in\mathbb{Z})$, leading to the disappearance of even-order responses.

In the main text and the rest of the SM, 
we have used the above argument for the selection rules of the HG spectra. 

\subsection{Laser-driven dynamics of electric polarizations under a static magnetic field}
\begin{figure}[t]
\begin{center}
\includegraphics[width=\columnwidth]{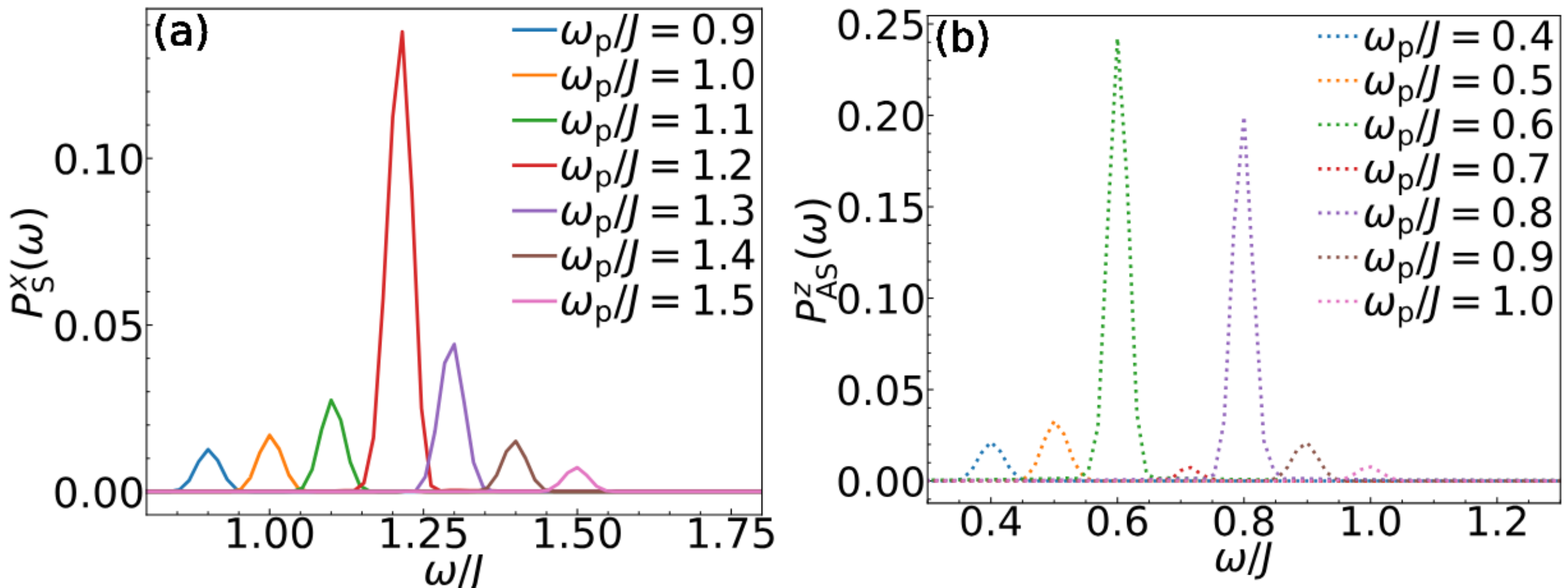}
\caption{
(a) $P_\mathrm{S}^x(\omega)$ and (b) $P_\mathrm{AS}^z(\omega)$ calculated in the model with $h_z/J=0.1$. 
We take several laser frequencies $\omega_{\rm p}$ around the triplon or triplon-pair resonance points. 
%The meaning of the vertical dotted line is the same as in Fig.~\ref{fig:P_h0_DSP_SS}. 
}
\label{fig:P_h_DSP_SS}
\end{center}
\end{figure}

In Fig.~2 of the main text, we have discussed the HG spectra of $\bm{P}_\mathrm{S}(\omega)$ and
$\bm{P}_\mathrm{AS}(\omega)$ in the absence of a DC magnetic field. 
Here, we compute the fundamental harmonics (linear response) of $P^x_\mathrm{S}(\omega)$ and
$P^z_\mathrm{AS}(\omega)$  under a magnetic field of $h_z=0.1J$, which are shown in Fig.~\ref{fig:P_h_DSP_SS}. 
In these figures, we adopt smaller light-matter coupling constants $|\Pi_{\rm me}E_{\rm p}|=0.01J$ and $|D_{\rm me}E_{\rm p}|=0.01J$ because we emphasize only the linear response here. 

As we mentioned in the main text, the $S=1$ ($S=0$) mode is known to be activated through the ME coupling of $\hat{\bm{P}}_\mathrm{AS}$ ($\hat{\bm{P}}_\mathrm{S}$)~\cite{Miyahara2023}. 
If this is true, the peak position of $\bm{P}_\mathrm{AS}(\omega)$ undergoes the Zeeman splitting, but $\bm{P}_\mathrm{S}(\omega)$ does not. 
As expected, Fig.~\ref{fig:P_h_DSP_SS} (a) shows that $P_\mathrm{S}^x(\omega)$ exhibits a single peak around $\omega_{\rm p}/J=1.2$, whose position is the same as the case without a DC magnetic field [see Fig. 2(a) in the main text]. 
Namely, ${\bm P}_\mathrm{S}(\omega)$ is not activated when $\omega_p$ deviates from the triplon-pair resonance frequency $1.2J$. 
On the other hand, Fig.~\ref{fig:P_h_DSP_SS} (b) shows that the resonance frequency of $P_\mathrm{AS}^z(\omega)$ is shifted and the intensities at $\omega_{\rm p}/J=0.6$ and $0.8$ are larger than that for the case of ${\bm H}={\bm 0}$ at $\omega_{\rm p}/J=0.7$ that is the resonance frequency of $\bm{P}_\mathrm{AS}(\omega)$ at $h_z=0$ [see Fig. 2(b) in the main text]. 
These three peaks would respectively correspond to the $S^z=1,~-1$ and 0 modes. 
This Zeeman splitting of the linear response of $P_\mathrm{AS}^z(\omega)$ under a DC magnetic field $h_z$ 
is also consistent with the previous study for a larger system~\cite{Miyahara2023}.
%As is discussed in the previous study~\cite{Miyahara2023}, we confirm that the $S^z=\pm 1~(0)$ triplon can be captured by $P_\mathrm{AS}^{z(x)}(\omega)$. 
%%%%%%%%%%%%%%%%%%%%%%%%%%%%%%%%%%%%%%%
%  fig. 3  %%
%%%%%%%%

%%%%%%%%%%%%%%%%%%%%%%%%%%%%%%%%%%%%%%%%%%%%
%%%%%%%%%%%%%%%%%%%%%%%%%%%%%%%%%%%%%%%%%%%%
%%%%%%%%%%%%%%%%%%%%%%%%%%%%%%%%%%%%%%%%%%%%
\subsection{Magnetization dynamics in the dimer singlet phase}
%Anisotropic interactions breaking the U(1) symmetry mix $S^z=\pm 1$ triplon excitation with $S^z=0$ singlet ground state and allow for the magnetic from the ground state to the excited one~\cite{Sakai2000}. 
%Therefore, we can examine this transition in the U(1) breaking case, i.e., including the intradimer DM interactions. 
%Throughout this Supplemental Material, we fix  $D'/J=0.015$ and $D/J=0.025$ for DM terms as in the main text. 
The main text has concentrated on the dynamics of the electric polarization in the SSM. On the other hand, 
in this section, we discuss the magnetization dynamics driven by the AC Zeeman interaction $\hat{\mathscr{H}}_\mathrm{Zeeman}(t)=-\sum_i\bm{h}_\mathrm{p}(t)\cdot\hat{\bm{S}}_i$. 
If we consider the SU(2)-symmetric model without DM interactions, the ground state of the dimer single phase (DSP) is a purely spin singlet state with $S=0$ and $S^z=0$. In the SU(2) case, the AC Zeeman interaction cannot activate the $S=1$ triplon excitations because the total spin is not changed by acting ${\hat S}^\alpha$ on the $S=0$ ground state.
On the other hand, if the system possesses DM interactions with $D$ and $D'$ like SCBO, the SU(2) symmetry is no longer preserved, and the ground-state wave function contains $S=1$ components. 
%Then, we can usually observe the triplon excitations through the AC Zeeman interaction.  
%where 
Therefore, in the U(1) breaking case with $D$ and $D'$, the triplons with $S=1$ (resonance frequency $\omega/J=0.7$) are expected to be active to the AC Zeeman term~\cite{Sakai2000} like the case of the ME coupling 
$\hat{\mathscr{H}}_\mathrm{ME}(t)=-{\bm E}_{\rm p}(t)\cdot{\hat{\bm P}}_\mathrm{AS}$. 
%The specific values about the DM interactions used in our calculation are $D/J=0.025$ for intradimer terms, and $D'/J=0.015$ for interdimer ones.

\begin{figure}[!h]
\begin{tabular}{cc}
	\begin{minipage}[c]{0.48\hsize}
	\centering
	\includegraphics[width=\columnwidth]{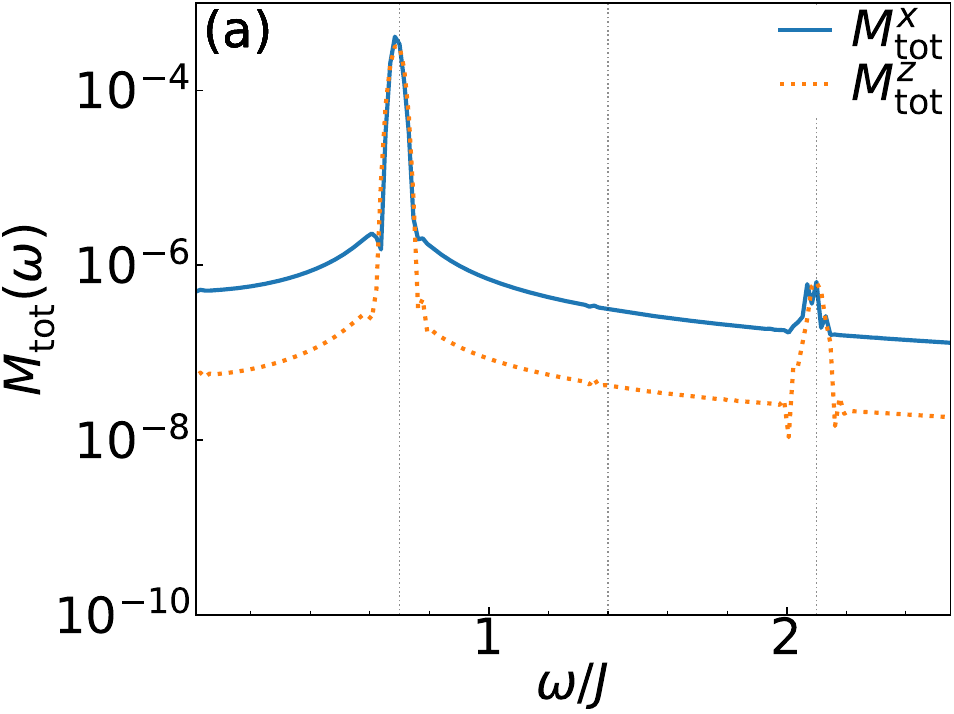}
	\end{minipage}
	\begin{minipage}[c]{0.48\hsize}
	\centering
	\includegraphics[width=\columnwidth]{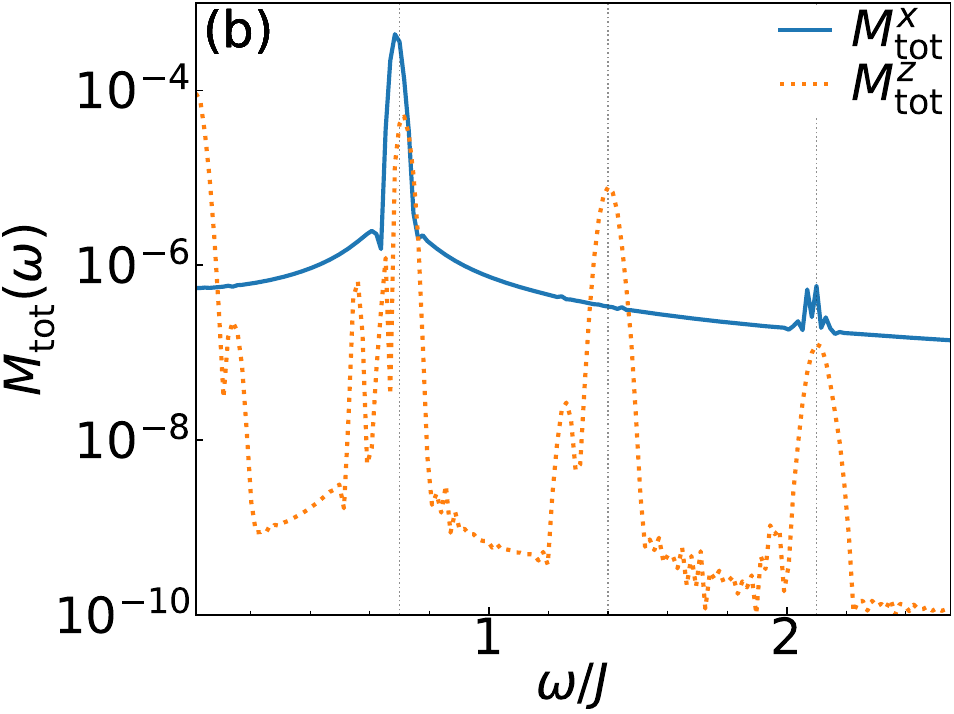}
	\end{minipage}
	%\begin{minipage}[c]{0.33\hsize}
	%\centering
	%\includegraphics[width=\columnwidth]{STO_Jnnn0.500_lnn1.000_lnnn1.000_D0.025_Dn0.000_Dns0.000_Dz0.%015_hx0.100_hz0.000_bias0.000000_h0.200000_w0.700_cyc400.pdf}
	%\end{minipage}
\end{tabular}
\caption{
HG spectra of $\bm{M}(\omega)$ in the U(1)-breaking SSM with an AC Zeeman interaction $-\sum_ih^\alpha_{\rm p}(t)\hat{S}_i^\alpha$ (a) under zero DC magnetic field and (b) under a finite DC magnetic field along the $z$ axis ($h_z/J=0.1$). 
The blue and orange lines in both figures represent the $x$ and $z$ components of the magnetization, respectively.
Vertical dotted lines represent integer multiples of $\omega_\mathrm{p}$. 
%The meaning of the vertical dotted line is the same as that in Fig.2~\ref{fig:P_h0_DSP_SS} in the main text. 
}
\label{fig:STO_DSP_SS}
\end{figure}

Figure~\ref{fig:STO_DSP_SS} (a) shows $M^\alpha(\omega)$ ($\alpha=x,~z$) driven by the AC Zeeman interaction $\hat{\mathscr{H}}_\mathrm{Zeeman}(t)=-\sum_i\bm{h}_\mathrm{p}(t)\cdot\hat{\bm{S}}_i$, with resonance frequency in zero DC magnetic field. 
Like electric polarizations (Fig.2 %~\ref{fig:P_h0_DSP_SS}
in the main text), third-order harmonics are also observed in addition to the fundamental harmonic, but even-order responses do not arise because of dynamical symmetries $\hat{U}\hat{\mathscr{H}}(t)\hat{U}^{-1}=\hat{\mathscr{H}}(t+T_\mathrm{p}/2)$ associated with $\hat{U}=C_2(z)$ and $\sigma_{zx}$ for $H_{\rm p}^x$, and $\hat{U}=\sigma_{zx}$ or $\sigma_{yz}$ for $H_{\rm p}^z$. 
%With the DC magnetic field, the Zeeman split of $\bm{M}(\omega)$ is confirmed from Fig.~\ref{fig:STO_DSP_SS} (b), which is same as $\bm{P}_\mathrm{AS}(\omega)$ under the DC magnetic field (Fig.3~\ref{fig:P_h_DSP_SS} in the main text). 
%$M^z$ and $M^x$ split under the DC magnetic field in the $z$ direction and the $x$ direction, respectively. 
%As for the magnetization, dynamical symmetries are easily broken when the DC magnetic field is in the same direction as the oscillating magnetic field, so that the even-order harmonics occur. 
We, however, find that the dynamical symmetries are easily broken when a DC magnetic field is added along the same direction as the oscillating magnetic field, without considering the symmetry of the system.
This is because the operator $\hat{U}$ should satisfy $\hat{U}\hat{S}^\alpha \hat{U}^\dag=-\hat{S}^\alpha$ for the conservation of the dynamical symmetry [see Eq.~\eqref{eq:quantity_SS}], but this relation changes the sign of a DC Zeeman interaction $\hat{\mathscr{H}}_\mathrm{Zeeman}=-\sum_ih_\alpha\cdot\hat{S}_i^\alpha$. 
Figure~\ref{fig:STO_DSP_SS} (b) is the spectra $M^\alpha(\omega)$ ($\alpha=x,~z$) under the DC magnetic field in the $z$ direction ($h_z/J=0.1$), so that even-order harmonics appear only in $M^z(\omega)$ but not in $M^x(\omega)$. 
In the case of the DC magnetic field along the $x$ direction, similarly to Fig.~\ref{fig:STO_DSP_SS} (b), only the dynamical symmetries for $M^x(\omega)$ are broken. 
More detail about the dynamical symmetries for the magnetization dynamics is discussed in the final section of this SM. 

%  fig. S1  %%
%%%%%%%%

%%%%%%%%%%%%%%%%%%%%%%%%%%%%%%%%%%%%
%%%%%%%%%%%%%%%%%%%%%%%%%%%%%%%%%%%%
%%%%%%%%%%%%%%%%%%%%%%%%%%%%%%%%%%%%
\subsection{Two-color laser}\label{sec:BCL4R_SS}
In the main text, we show the HG spectrum driven by a two-color laser with 3-fold rotational symmetry and find that the SSM does not have this symmetry, resulting in the appearance of even-order responses.
Contrarily, in this section, we show that the dynamical symmetry exists even under the two-color laser by properly tuning their frequencies. 

\begin{figure}[!h]
\begin{tabular}{cc}
        \vspace{5pt}
        \hspace{10pt}
	\begin{minipage}[c]{0.38\hsize}
	\centering
	\includegraphics[width=\columnwidth]{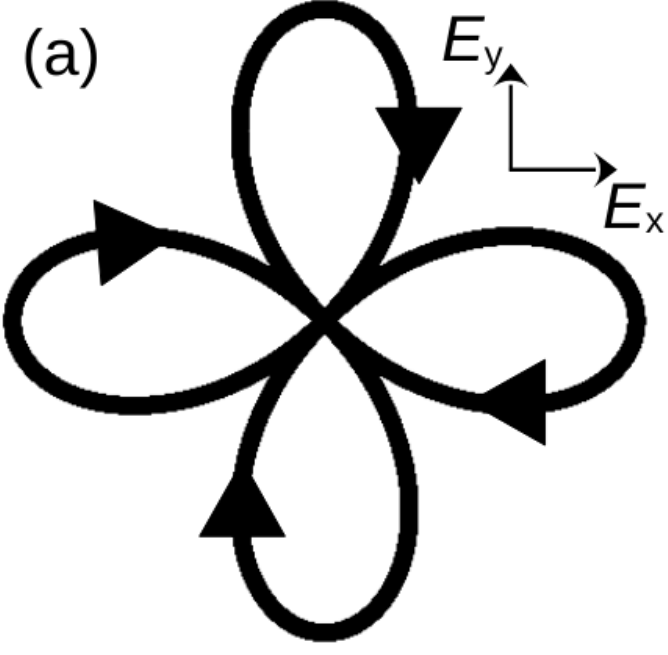}
	\end{minipage}
        \hspace{10pt}
	\begin{minipage}[c]{0.48\hsize}
	\centering
	\includegraphics[width=\columnwidth]{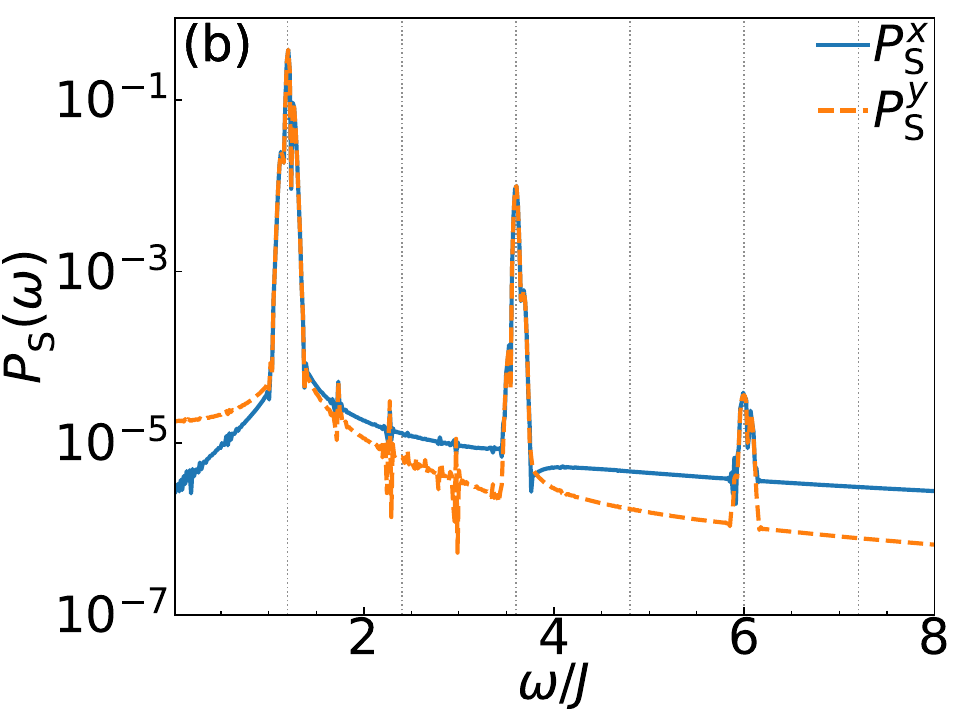}
	\end{minipage}\\
 	\begin{minipage}[c]{0.48\hsize}
	\centering
	\includegraphics[width=\columnwidth]{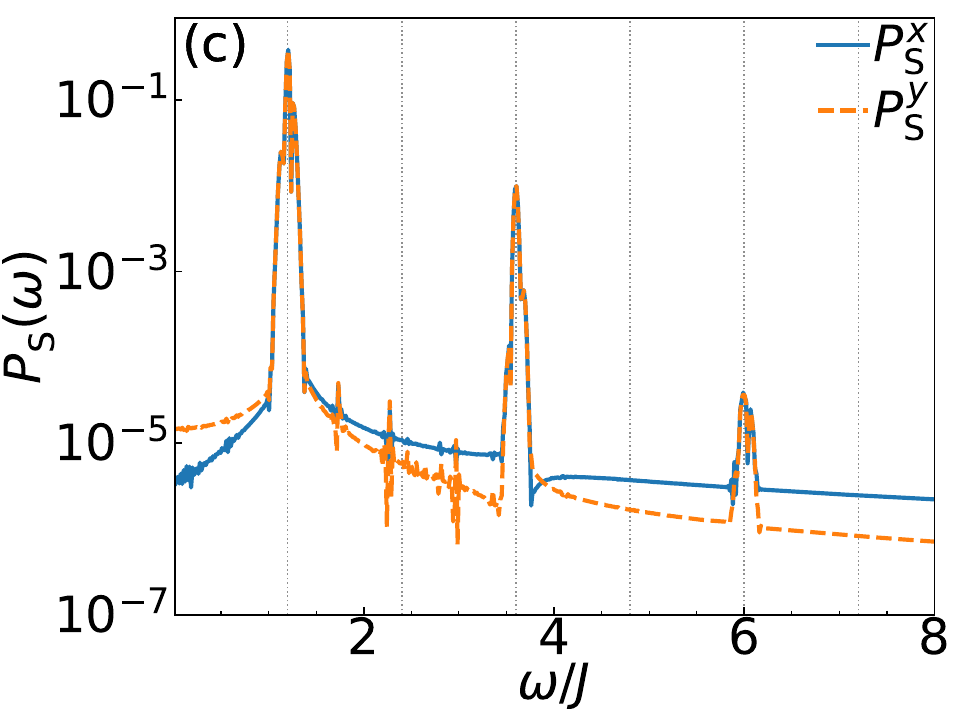}
	\end{minipage}
	\begin{minipage}[c]{0.48\hsize}
	\centering
	\includegraphics[width=\columnwidth]{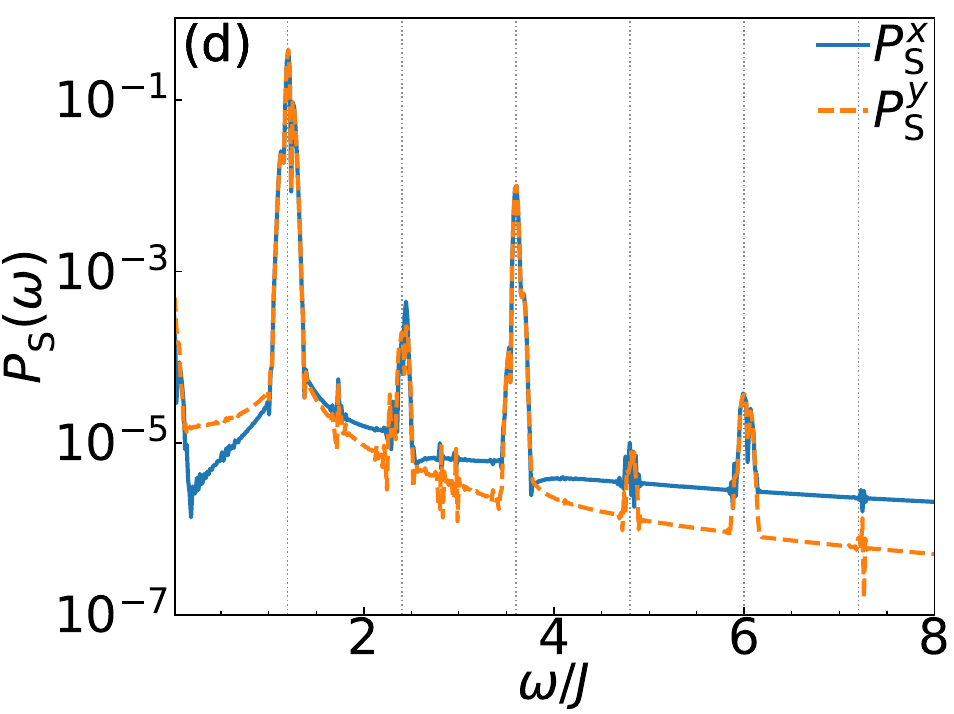}
	\end{minipage}
 \end{tabular}
\caption{
(a) Trajectory of the laser field of bicircular light [Eq.(5)%~(\ref{eq:bicircular}
) of the main text in $\ell=3$] with the 4-fold rotational symmetry. 
%(b) The electric polarization $\bm{P}_\mathrm{S}(\omega)$ under the bicircular light at $D=0$, $D'/J=0.015$ without a DC magnetic field.
Harmonic spectra $\bm{P}_\mathrm{S}(\omega)$ in the U(1)-breaking SSM irradiated by the ME coupling $\hat{\mathscr{H}}_\mathrm{ME}(t)=-E^\alpha_{\rm p}(t)\cdot{\hat P}^\alpha_\mathrm{S}$ of the bicircular light (b) in zero DC magnetic field, in a DC magnetic field along (c) the $z$ axis ($h_z/J=0.1$) and (d) the $x$ axis ($h_x/J=0.1$).
The blue and orange lines, respectively, represent $P_\mathrm{S}^x(\omega)$ and $P_\mathrm{S}^y(\omega)$.
Vertical dotted lines represent integer multiples of $\omega_\mathrm{p}$. 
%The meaning of the vertical dotted line is the same as that in Fig.2~\ref{fig:P_h0_DSP_SS} in the main text. 
}
\label{fig:PSS_circularly_DSP_SS}
\end{figure}

We choose $\ell=3$ in Eq.(5) %~(\ref{eq:bicircular}
of the main text, in which the trajectory of the laser field satisfies the 4-fold rotational symmetry as shown in Figure~\ref{fig:PSS_circularly_DSP_SS} (a). 
In this case, both the SSM and two-color laser obey the symmetry of $\pi$ rotation around the $z$ axis [$C_2(z)$]. 
%and the two-fold symmetry is included in 4-fold symmetry.
Figure~\ref{fig:PSS_circularly_DSP_SS} (b) shows the spectra $P^{x,y}_\mathrm{S}(\omega)$ under this bicircular light with $\ell=3$ in the U(1) breaking case with $D/J=0.025$, $D'/J=0.015$ and ${\bm H}={\bm 0}$. 
We find that odd-order peaks appear, but even-order responses are still inactive even in the low-symmetric U(1) breaking case.

%According to the Upper Table ~\ref{tb:symmetry_SS}, the SSM in the DSP possesses the $C_2(z)$ symmetry, and the dynamical symmetry for $P^{x,y}_\mathrm{S}(\omega)$ leads to the absence of even responses. 
The absence of even-order responses stems from the dynamical symmetry associated with $C_2(z)$ (for more detail, see the upper Table~\ref{tb:symmetry_SS}). 
%The fact that $C_2(z)$ symmetry is associated with the two-color laser with the four-fold rotational symmetry is confirmed by applying a DC magnetic field.
However, one sees that the symmetry $C_2(z)$ remains under a DC magnetic field along the $z$ axis, while it is destroyed if a DC magnetic field is parallel to the $x$ or $y$ axis.
Figure~\ref{fig:PSS_circularly_DSP_SS} (c) shows the spectra under a finite DC magnetic field along the $z$ axis ($h_z/J=0.1$) in the same model 
 as Fig.~\ref{fig:PSS_circularly_DSP_SS} (a) except for the magnetic field.
In this case, the $C_2(z)$ symmetry survives, and the selection rule is unchanged; therefore, we find the absence of the even-order responses.
On the other hand, under a DC magnetic field along the $x$ axis ($h_x/J=0.1$), the even-order peaks appear as shown in Fig.~\ref{fig:PSS_circularly_DSP_SS} (d), due to the $C_2(z)$ symmetry breakdown.
These results indicate that tuning the direction of a DC magnetic field can control the breakdown or conservation of the symmetry relevant to the selection rule in the present SSM.

%  fig. S2  %%
%%%%%%%%

%%%%%%%%%%%%%%%%%%%%%%%%%%%%%%%%%%%%
%%%%%%%%%%%%%%%%%%%%%%%%%%%%%%%%%%%%
%%%%%%%%%%%%%%%%%%%%%%%%%%%%%%%%%%%%
\subsection{Dynamical symmetries in the empty plaquette phase}\label{sec:EPP_SS}
As we mentioned in the Introduction of the main text, if we introduce a strong enough perturbation like a pressure or a magnetic field, SCBO enters in other phases accompanying a spontaneous symmetry breaking. 
%Here, we analyze the harmonic spectra in two ordered phases, the empty plaquette phase (EPP) and 1/2 plateau phase (PP) [see Fig.~\ref{fig:model_SS} (c) and (d)]. 
In the main text, we discuss the harmonic spectra in the 1/2 plateau phase (PP) [see Fig.1 %~\ref{fig:model_SS}
(d) in the main text]. 

\begin{figure}[!h]
\begin{tabular}{cc}
        \vspace{5pt}
        \hspace{10pt}
	\begin{minipage}[c]{0.38\hsize}
	\centering
	\includegraphics[width=\columnwidth]{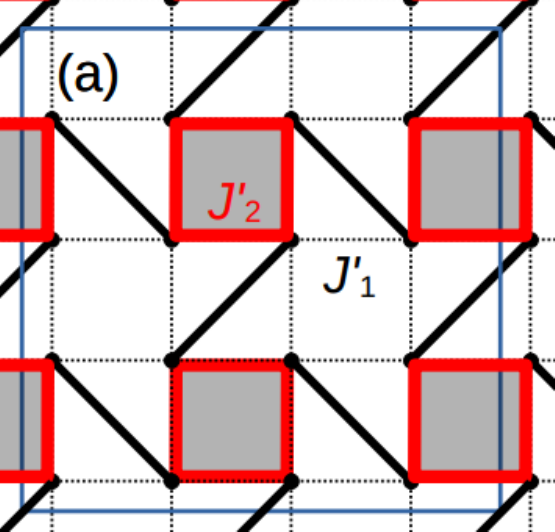}
	\end{minipage}
        \hspace{10pt}
	\begin{minipage}[c]{0.48\hsize}
	\centering
	\includegraphics[width=\columnwidth]{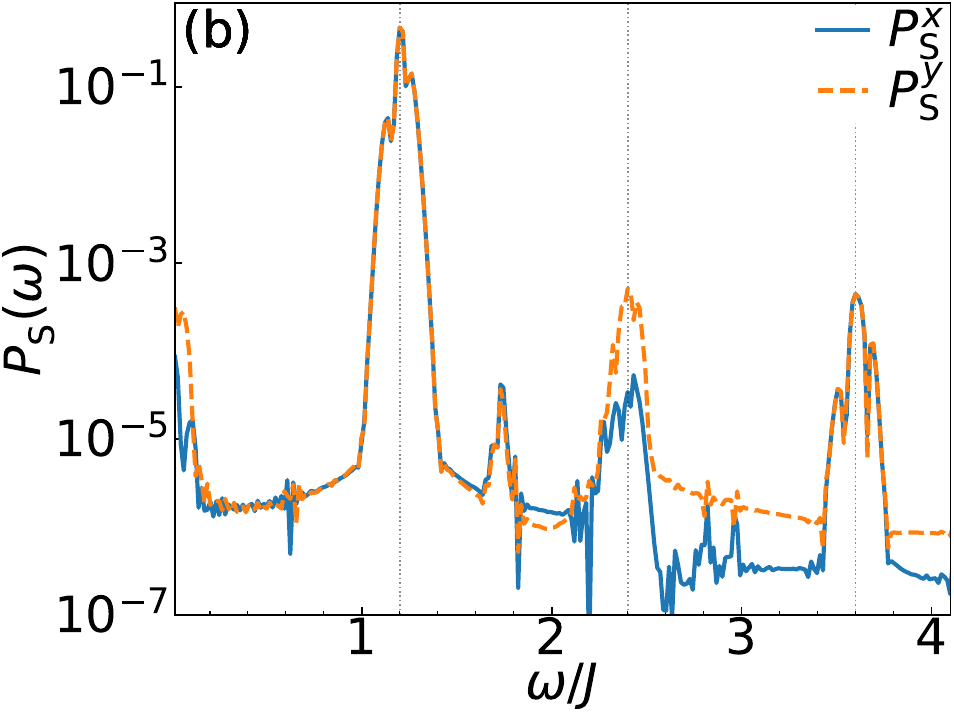}
        \end{minipage}\\
	\begin{minipage}[c]{0.48\hsize}
	\centering
	\includegraphics[width=\columnwidth]{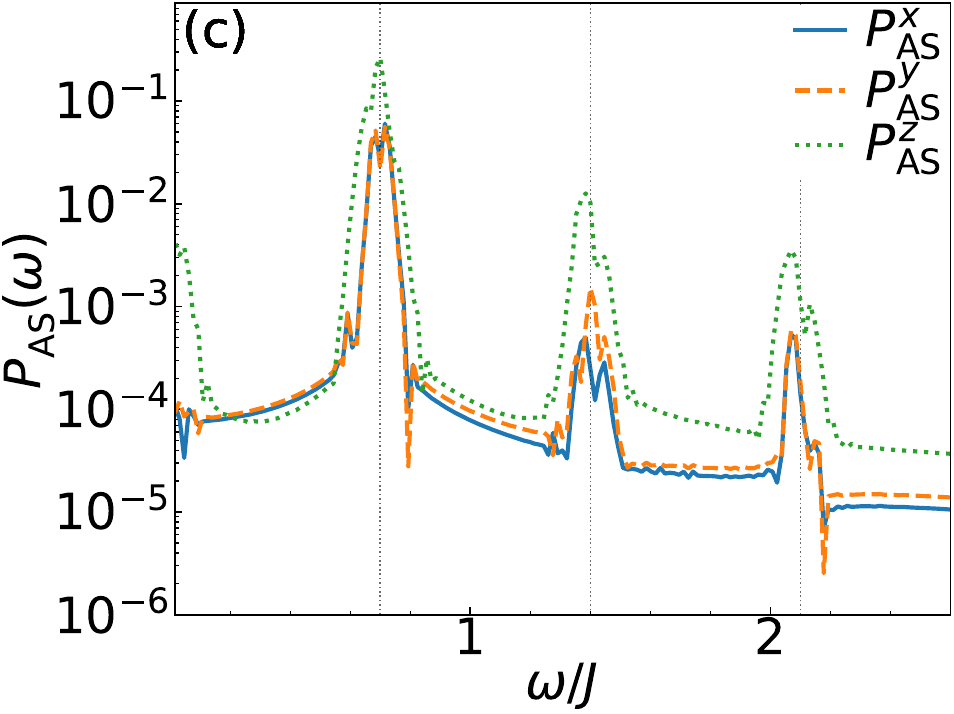}
	\end{minipage}
	\begin{minipage}[c]{0.48\hsize}
	\centering
	\includegraphics[width=\columnwidth]{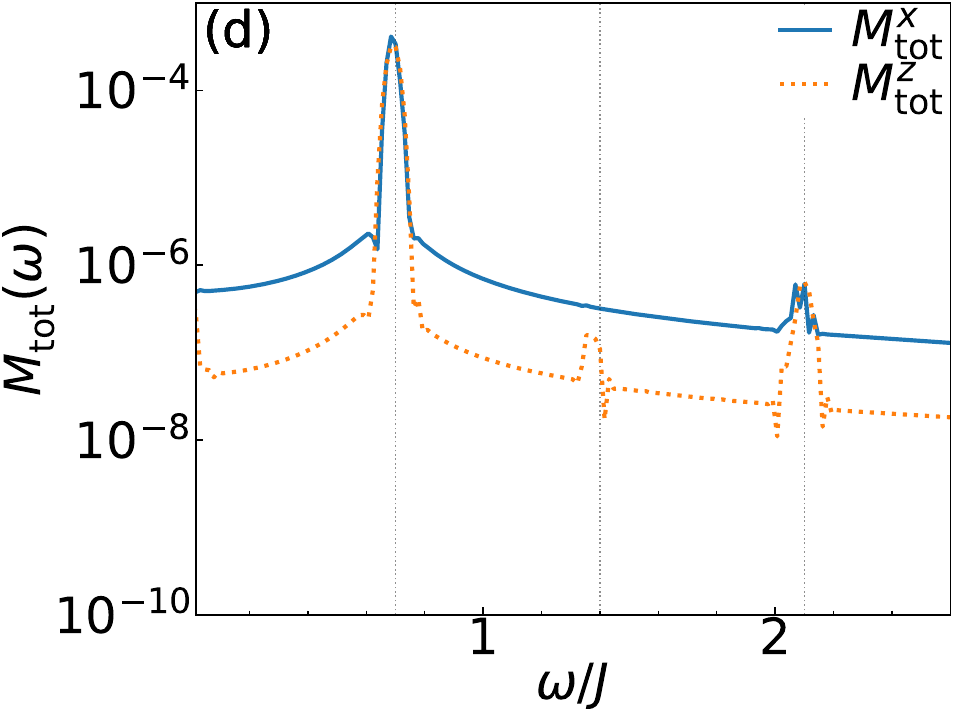}
        \end{minipage}
\end{tabular}
\caption{
(a) Ordering pattern of the EPP (four spins on each red square form a singlet), in which we introduce exchange interactions $J'_1$ (dotted black lines) and $J'_2$ (bold red lines) to stabilize the EPP~\cite{Takushima2001,Boos2019}. 
The shaded area represents the singlet state. 
The diagonal bonds indicated with bold black lines denote the intradimer interaction $J$.
The blue line encloses the $16$-site square cluster used in our calculation. 
ME-coupling driven harmonic spectra (b) $\bm{P}_\mathrm{S}(\omega)$ and (c) $\bm{P}_\mathrm{AS}(\omega)$ in the EPP of the SSM with $\lambda'=0.01$ and a DC magnetic field along the $x$ axis ($h_x/J=0.1$). 
Compared with the result in the DSP [Fig. 3 %~\ref{fig:P_DSP_SS}
(a) for $\bm{P}_\mathrm{S}(\omega)$ and Fig. 3 %~\ref{fig:P_DSP_SS}
(b) for $\bm{P}_\mathrm{AS}(\omega)$], even-order peaks of $P_\mathrm{S}^x(\omega)$ and $P_\mathrm{AS}^x(\omega)$, which do not exist in the DSP, are observed in the EPP. 
The blue and orange lines in both panels respectively represent the $x$ and $y$ components of the two electric polarizations, respectively, while the green line in (c) shows the $z$ component of $\bm{P}_\mathrm{AS}(\omega)$.
%The meaning of the vertical dotted line is the same as in Fig.~\ref{fig:P_h0_DSP_SS}. 
(d) AC-Zeeman-term driven spectrum $\bm{M}(\omega)$ in the EPP of the SSM with $\lambda'=0.01$ and zero DC magnetic field. 
Compared with the result of the DSP [Fig.~\ref{fig:STO_DSP_SS} (a)], even-order peaks of $M^z(\omega)$, which do not exist in the DSP, are observed in the EPP. 
The blue and orange lines represent the $x$ and $z$ components of the magnetization.
Vertical dotted lines represent integer multiples of $\omega_\mathrm{p}$. 
%The meaning of the vertical dotted line from (b) to (d) is the same as in Fig.2~\ref{fig:P_h0_DSP_SS} in the main text. 
}
\label{fig:EPP_SS}
\end{figure}

In this section, we calculate harmonic spectra in another phase, the empty plaquette phase (EPP). % [see Fig.1~\ref{fig:model_SS} (c) in the main text]. 
%In this section, we discuss the detail of EPP and the results of dynamical symmetries in this phase.
It is known that two quantum phase transitions occur and three phases appear in the SSM on the line of $J'/J$: the DSP with a finite spin gap, the plaquette state in the intermediate phase, and the gapless antiferromagnetic phase (AFM)~\cite{Koga2000,Corboz2013}. See Fig.1 %~\ref{fig:model_SS}
(c) in the main text. 
The intermediate plaquette state in the SSM is the ordered state in which the four spins on the square without diagonal bonds form a singlet as a unit, shown in Fig.~\ref{fig:EPP_SS} (a), and then it is called the empty plaquette phase (EPP). 
The mirror symmetries $\sigma_{zx}$ and $\sigma_{yz}$ are spontaneously broken in the EPP. 
%In extending the interactions of the interdimer interactions $J'$ in the SSM to take two different values, $J'_1$ and $J'_2$, as shown in Fig.~\ref{fig:EPP_SS} (a), the EPP becomes stable when $J'_1/J'_2\gg 1$ or $J'_1/J'_2\ll 1$~\cite{Koga2000}. 
To realize the symmetry-breaking pattern of the EPP in our 16-site cluster, 
%we introduce small additional exchange interactions $\lambda'J'$ along the red bonds in Fig.1%~\ref{fig:model_SS}
%(c) in the main text [see also Fig.~\ref{fig:EPP_SS} (a)]. 
%Since our purpose is just to reduce the symmetry from the DSP, we adopt perturbations not in the extreme, 
we introduce the order parameter of this phase as a small perturbation term $\hat{\mathscr{H}}_{\lambda'}=J'\lambda'\sum\left(\hat{\bm{S}}_i\cdot\hat{\bm{S}}_j-\hat{\bm{S}}_i\cdot\hat{\bm{S}}_{j'}\right)$. 
The sum of $ij$ ($ij'$) is taken over red (black) squares in Fig.~\ref{fig:EPP_SS} (a). 
As a result, the system possesses two sorts of exchange interactions $J'_1=J'(1-\lambda')$ and $J'_2=J'(1+\lambda')$ shown in Fig.~\ref{fig:EPP_SS} (a), instead of the original $J'$. 
We set $\lambda'=0.01$ ($\lambda'=0$ in the DSP) in the numerical calculation. 

%Firstly, we show the result in the EPP, which resides between the SDP and N\'eel phase, as shown in Fig.~\ref{fig:model_SS} (c). 
Firstly, we compare the  harmonic spectra for $\bm{P}_\mathrm{S(AS)}(\omega)$ in the DSP and EPP with the ME coupling $\hat{\mathscr{H}}_\mathrm{ME}(t)=-E^\alpha_{\rm p}(t)\cdot{\hat P}^\alpha_\mathrm{S(AS)}$.
If we apply a magnetic field $h_x$ in the DSP, among the spacial symmetries, only the $\sigma_{yz}$ symmetry survives, which results in the absence of even-order peaks of $P_\mathrm{S}^x(\omega)$ and $P_\mathrm{AS}^x(\omega)$, as shown in  Fig.3 %~\ref{fig:P_DSP_SS}
(a) and (b) in the main text. 
On the other hand, (as we mentioned) the mirror symmetries $\sigma_{zx}$ and $\sigma_{yz}$ are spontaneously broken in the EPP even without the application of a DC magnetic field. 
%so that the dynamical symmetries of $P_\mathrm{S}^x(\omega)$ and $P_\mathrm{AS}^x(\omega)$ under the DC magnetic field in the $x$ direction are destroyed. 
Figure~\ref{fig:EPP_SS} (b) and (c) indeed show that all components of electric polarizations $P^\alpha_\mathrm{S}(\omega)$ and $P^\alpha_\mathrm{AS}(\omega)$ ($\alpha=x,~y,~z$) %and we find that all component including $\alpha=x$
%the intradimer DM interactions ($D/J=0.025$ and $D'/J=0.015$) and 
have even-order peaks in the EPP with $h_x/J=0.1$. 
%We confirm even-order responses for these physical quantities, for which dynamical symmetries are broken. 
%We discuss more detail about EPP in Section 3~\ref{sec:EPP_SS} of the SM~\cite{SM}. 

Next, we consider the magnetization dynamics driven by an AC Zeeman interaction in the EPP. 
For the DSP driven by an AC Zeeman term, 
we find dynamical symmetries of $\hat{U}\hat{\mathscr{H}}(t)\hat{U}^{-1}=\hat{\mathscr{H}}(t+T_\mathrm{p}/2)$ when the total Hamiltonian is given by 
$\hat{\mathscr{H}}(t)=\hat{\mathscr{H}}_\mathrm{SS}+\hat{\mathscr{H}}_\mathrm{DM}+\hat{\mathscr{H}}_\mathrm{Zeeman}(t)$; the symmetry operator $\hat{U}$ is chosen to be $C_2(z)$ and $\sigma_{zx}$ for $H_{\rm p}^x$, $\sigma_{zx}$ and $\sigma_{yz}$ for $H_{\rm p}^z$.
However, we notice that the symmetries $\sigma_{zx}$ and $\sigma_{yz}$, which exist in the DSP with finite intradimer DM interactions, are spontaneously broken in the EPP.
This breakdown of the dynamical symmetry leads to the appearance of both even- and odd-order responses in $M^z(\omega)$ regardless of a DC field, as shown in Fig.~\ref{fig:EPP_SS} (d) (for more detail, see the Upper Table~\ref{tb:symmetry_SS}). 
%The parameters for the DM interactions of this figure are the same as Fig.~\ref{fig:STO_DSP_SS}.
%Even-order responses for $M^z(\omega)$, whose dynamical symmetries no longer exist, are confirmed. 
We stress that the irrelevance of the DC field in the spectrum of $M^z(\omega)$ is in contrast to the case of polarizations $P_\mathrm{S}^y(\omega)$ and $P_\mathrm{AS}^{y,z}(\omega)$, depicted in Fig.~\ref{fig:EPP_SS} (b) and (c), in which 
a finite DC field breaks the dynamical symmetries $\hat{U}\hat{\mathscr{H}}(t)\hat{U}^{-1}=\hat{\mathscr{H}}(t+T_\mathrm{p}/2)$ if the symmetry operator $\hat{U}$ is chosen to be $C_2(z)$ or $\sigma_{yz}$ for $E_{\rm p}^x$, $C_2(z)$ or $\sigma_{zx}$ for $E_{\rm p}^y$, and $S_4$ for $E_{\rm p}^z$. 

%  fig. S3  %%
%%%%%%%%

%%%%%%%%%%%%%%%%%%%%%%%%%%%%%%%%%%%%
%%%%%%%%%%%%%%%%%%%%%%%%%%%%%%%%%%%%
%%%%%%%%%%%%%%%%%%%%%%%%%%%%%%%%%%%%
\subsection{Dynamical symmetries in the full plaquette phase}
\begin{figure}[!h]
\begin{tabular}{cc}
	\begin{minipage}[c]{0.38\hsize}
	\centering
	\includegraphics[width=\columnwidth]{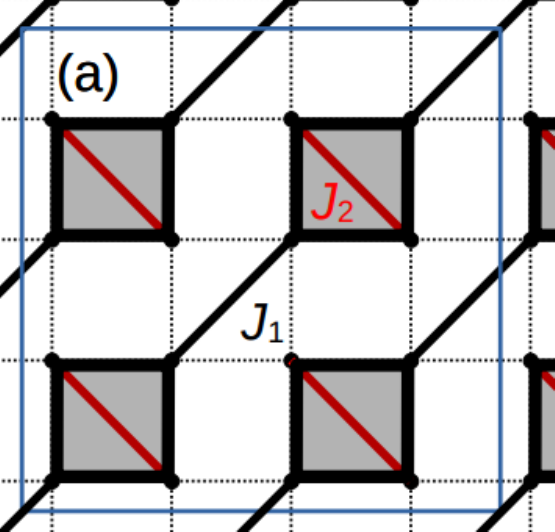}
	\end{minipage}
        \hspace{10pt}
	\begin{minipage}[c]{0.5\hsize}
	\centering
	\includegraphics[width=\columnwidth]{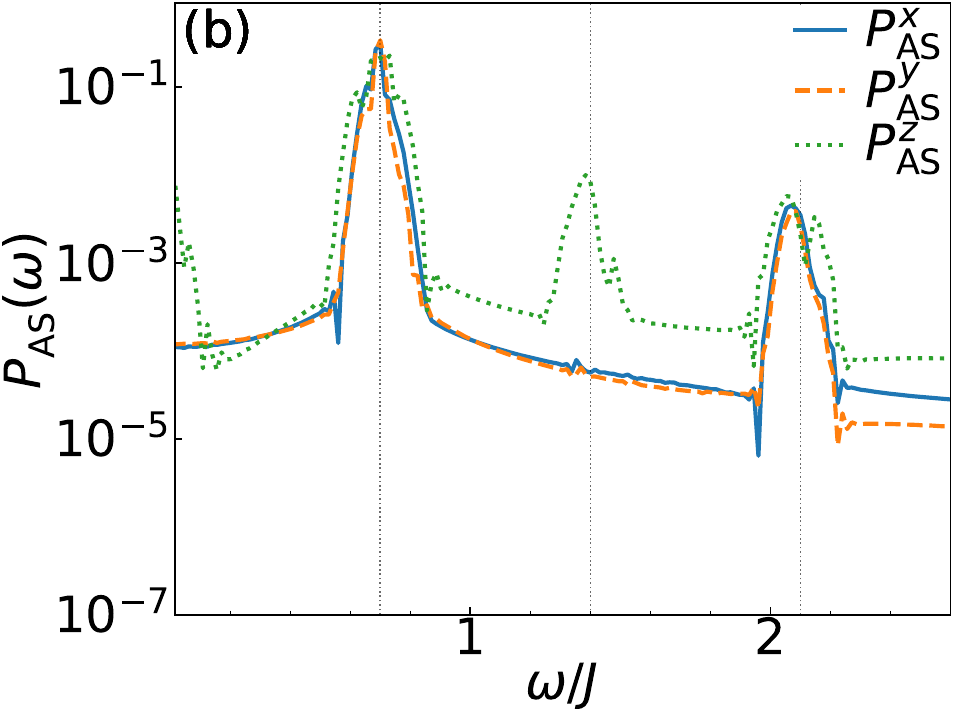}
	\end{minipage}
\end{tabular}
\caption{
(a) Ordering pattern of the FPP, containing two different diagonal bonds $J_1$ (bold black lines) and $J_2$ (bold red lines)~\cite{Takushima2001,Boos2019}. Four spins in each shaded square form a singlet. 
Both horizontal and vertical lines are interdimer interactions $J'$.
The blue line encloses the $16$-site square cluster used in our numerical calculation. 
(b) HG spectra of electric polarizations $\bm{P}_\mathrm{AS}(\omega)$ in the SSM with $\lambda=0.01$ driven by the ME coupling $\hat{\mathscr{H}}_\mathrm{ME}(t)=-E^\alpha_{\rm p}(t)\cdot{\hat P}^\alpha_\mathrm{AS}$. 
Even-order peaks of $P_\mathrm{AS}^z$, which do not exist in the DSP, emerge in the FPP. 
The blue, orange, and green lines respectively represent the $x$, $y$, and $z$ components of $\bm{P}_\mathrm{AS}(\omega)$.
Vertical dotted lines represent integer multiples of $\omega_\mathrm{p}$. 
}
\label{fig:PAS_FPP_SS}
\end{figure}

SCBO, under high pressure, is known to undergo the transition to another plaquette phase from the DSP. This plaquette phase differs from the EPP. 
In the EPP, each strong singlet state resides on an empty plaquette, while the plaquette phase of SCBO under high pressure has strong singlet squares on the plaquettes with diagonal $J$ bonds, as shown in Fig.~\ref{fig:PAS_FPP_SS} (a)~\cite{Zayed2017}. This phase is called the full plaquette phase (FPP). 

To describe the FPP, we assume that the diagonal $J$ bond takes two different values, $J_1$ and $J_2$, and set $J_1=J(1-\lambda)$ and $J_2=J(1+\lambda)$ (the DSP corresponds to $\lambda=0$). 
In other words, we introduce the order parameter of this phase to the Hamiltonian as a small perturbation term $\hat{\mathscr{H}}_{\lambda}=J\lambda\sum\left(\hat{\bm{S}}_i\cdot\hat{\bm{S}}_j-\hat{\bm{S}}_i\cdot\hat{\bm{S}}_{j'}\right)$. 
The sum of $ij$ ($ij'$) is taken over red (black) diagonal bonds in Fig.~\ref{fig:PAS_FPP_SS} (a).

The FPP takes two different exchange interactions not only on the diagonal $J$ bond but also on the square bond around it, as shown by the bold black lines in Fig.~\ref{fig:PAS_FPP_SS} (a). Therefore, it seems to be necessary to extend the interdimer interactions $J'$ to two different values [bold and thin black lines of Fig.~\ref{fig:PAS_FPP_SS} (a)] in addition to the intradimer interactions $J_{1,2}$. 
However, if we focus on the existence or absence of dynamical symmetries, it is enough to consider only the extension of intradimer interactions to discuss the essential nature of laser-driven harmonic spectra in the FPP. This is because the introduction of two intradimer bonds $J_{1,2}$ reduces the symmetry enough. Therefore, we numerically compute the harmonic spectrum for the FPP by introducing $J_{1,2}$ in our 16-site cluster. 

In the FPP, the rotoreflection symmetry $S_4$ is spontaneously broken because of the inequivalent two sorts of intradimer bonds. 
This symmetry breaking is the same as that in the $1/2$ plateau phase (PP), where the dynamical symmetry for the polarization $P_\mathrm{AS}^z(\omega)$ disappears due to the lack of the $S_4$ symmetry as discussed in the main text. Therefore, we expect that the same dynamical symmetry breaking also occurs in the FPP. In fact, from Fig.~\ref{fig:PAS_FPP_SS} (b), we can confirm that the even-order responses of $P_\mathrm{AS}^z(\omega)$ occur in the FPP, while there are no even-order responses of $P_\mathrm{AS}^{x,y}(\omega)$ because the dynamical symmetry associated with $C_2(z)$ (see Table~\ref{tb:symmetry_SS}) is still conserved in the FPP.

%  fig. S4  %%
%%%%%%%%

%%%%%%%%%%%%%%%%%%%%%%%%%%%%%%%%%%%%
%%%%%%%%%%%%%%%%%%%%%%%%%%%%%%%%%%%%
%%%%%%%%%%%%%%%%%%%%%%%%%%%%%%%%%%%%
\subsection{Dynamical symmetries in the plaquette spin-nematic phase}
\begin{figure}[!h]
\begin{center}
\includegraphics[width=\columnwidth]{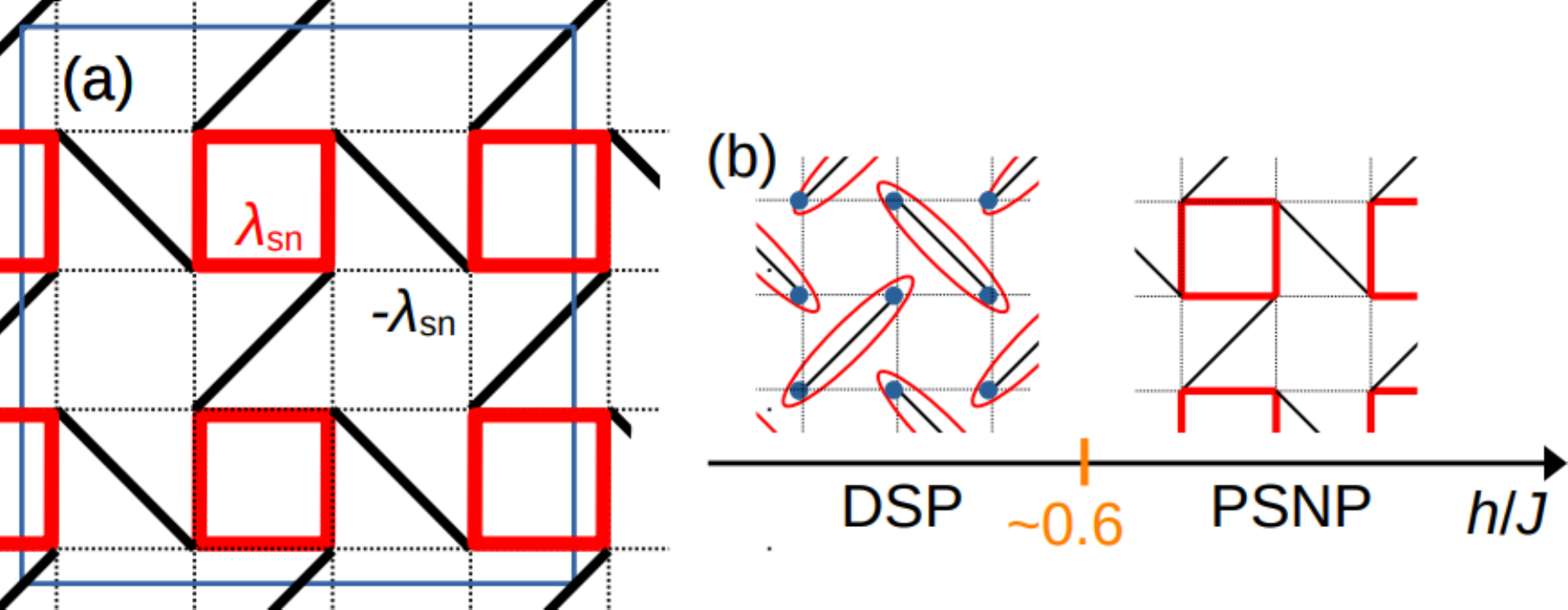}
\caption{
(a) Ordering pattern of the PSNP. In our numerical calculation, we introduce  small perturbation terms $\lambda_\mathrm{sn}(\hat{S}_i^+\hat{S}_j^++{\rm H.c.})$ on the bold red bonds and $-\lambda_\mathrm{sn}(\hat{S}_i^+\hat{S}_{j'}^++{\rm H.c.})$ on the dotted black bonds to stabilize the PSNP~\cite{Batista2018}. 
In other words, bold red lines stand for a positive spin-nematic bonds, while dotted black lines do a negative bonds. 
%The diagonal bonds indicated with bold black lines denote the intradimer interaction $J$.
The blue line encloses the $16$-site square cluster used in our calculation. 
(b) Ground-state phase diagram under the DC magnetic field along the $z$ axis for SU(2) symmetric model without DM terms $D'=0,~D=0$ at $J'/J=0.5$. 
The DSP and PSNP refer to the dimer singlet phase and the plaquette spin-nematic phase, respectively. 
Increasing the magnetic field, the PSNP emerges next to the DSP~\cite{Batista2018}. 
}
\label{fig:PSNP_SS}
\end{center}
\end{figure}

This section focuses on the triplon-pair condensed state, which is particularly interesting in the low-field region below the magnetization plateau phases.
Both theoretical~\cite{Momoi2000,Corboz2014,Schneider2016,Haravifard2016} and experimental~\cite{Imajo2022,Fogh2024} studies point out that the triplon pairs (two-triplon bound states) with $S^z=2$ first condense rather than triplons when the energy gap of triplon pair is closed by the application of a magnetic field. 
%Furthermore, it is theoretically shown that this $S^z=2$ triplon-pair condensed phase forms a pinwheel pattern with a singlet plaquette at the center, reminiscent of the EPP~\cite{Corboz2014}.
%The pinwheel pattern structure of the condensed phase breaks mirror symmetry but locally preserves rotational symmetries. 
%triplon bound states found in the dilute limit are the same ones~\cite{Batista2018}. 
When quasi particles with $S^z=2$ are condensed in quantum spin systems, a spin-nematic (quadrupolar-ordered) phase often appears~\cite{Chiral1991,Shannon2006,Ueda2009,Zhitomirsky2010,Penc2011,Sato2013-2,Hikihara2008,Sudan2009} 
%many papers are cited
and its order parameter is given by the expectation value of a tensor product of two spins like $\langle {\hat S}_j^+{\hat S}_k^+ +{\rm H.c.}\rangle$. In fact, the SU(2) symmetric SSM (without DM terms $D'=0,~D=0$) is predicted to have spin-nematic phases under a moderate magnetic field~\cite{Batista2018}: In the range $0.40\lesssim J'/J\lesssim 0.66$ that  covers both our adopted value ($J'/J=0.5$) and the value of SCBO ($J'/J\simeq 0.635$), the plaquette spin-nematic phase (PSNP) appears~\cite{Batista2018}. 
Figure~\ref{fig:PSNP_SS} (a) is the schematic view of this spin-nematic phase, where a positive order parameter resides on bold red bonds of each plaquette with four spins while a negative value resides on dotted black bonds.
In light of these previous studies, we here consider the HG spectra and the dynamical symmetries in the PSNP, which is theoretically predicted to appear next to the DSP in a low-magnetic-field region~\cite{Batista2018}, as shown in Fig.~\ref{fig:PSNP_SS} (b) for SU(2) symmetric SSM. A recent experimental study also implies the emergence of a spin-nematic like state in a low-field area~\cite{Imajo2022}. 

%  fig.   %%
%%%%%%%%

The PSNP spontaneously breaks the  U(1) spin-rotational symmetry around the $S^z$ axis in the SU(2) symmetric SSM (without DM terms $D'=0,~D=0$)~\cite{Momoi2000,Batista2018}, but DM terms exist in SCBO, extrinsically breaking both the SU(2) and U(1) symmetries. 
In the SSM without DM interactions, the phase transition to the PSNP breaks the spin-rotational symmetry spontaneously. 
In contrast to such a SU(2) model, the PSNP is expected to gradually change from the DSP when we apply a magnetic field to the U(1)-breaking SSM with finite DM terms. 
Nevertheless, compared with DSP, the mirror symmetries $\sigma_{zx}$ and $\sigma_{yz}$ are spontaneously broken in PSNP as in EPP. Therefore, we can distinguish the DSP and PSNP by using the mirror symmetries even when the SSM has DM terms.

%The main mode excited by $\hat{\bm{P}}_\mathrm{S}$ is the triplon bound state with $S=0$ (resonance frequency $\omega/J\sim 1.2$), while that by $\hat{\bm{P}}_\mathrm{AS}$ is the single triplon (resonance frequency $\omega/J\sim 0.7$)~\cite{Totsuka2001,Miyahara2023}. 
It is known~\cite{Nojiri2003,Miyahara2023} that the $(S,S^z)=(2,2)$ bound state of two triplons can be strongly excited by the ME coupling with $P_\mathrm{AS}^z(\omega)$ near the magnetic-field induced transition between the DSP and a magnetized PSNP. In the presence of such a moderate magnetic field, the energy gap of the $(S,S^z)=(2,2)$ triplon pair becomes small due to the Zeeman splitting, while the $(S,S^z)=(2,0)$ triplon pair has a large excitation gap regardless of the magnetic field. 
Below, we analyze the laser-driven dynamics of $\bm{P}_\mathrm{AS}(\omega)$ focusing on both the low-energy $(S,S^z)=(2,2)$ and the high-energy $(S,S^z)=(2,0)$ modes under the assumption that the PSNP emerges. 

\begin{figure}[!h]
\begin{tabular}{cc}
	\begin{minipage}[c]{0.48\hsize}
	\centering
	\includegraphics[width=\columnwidth]{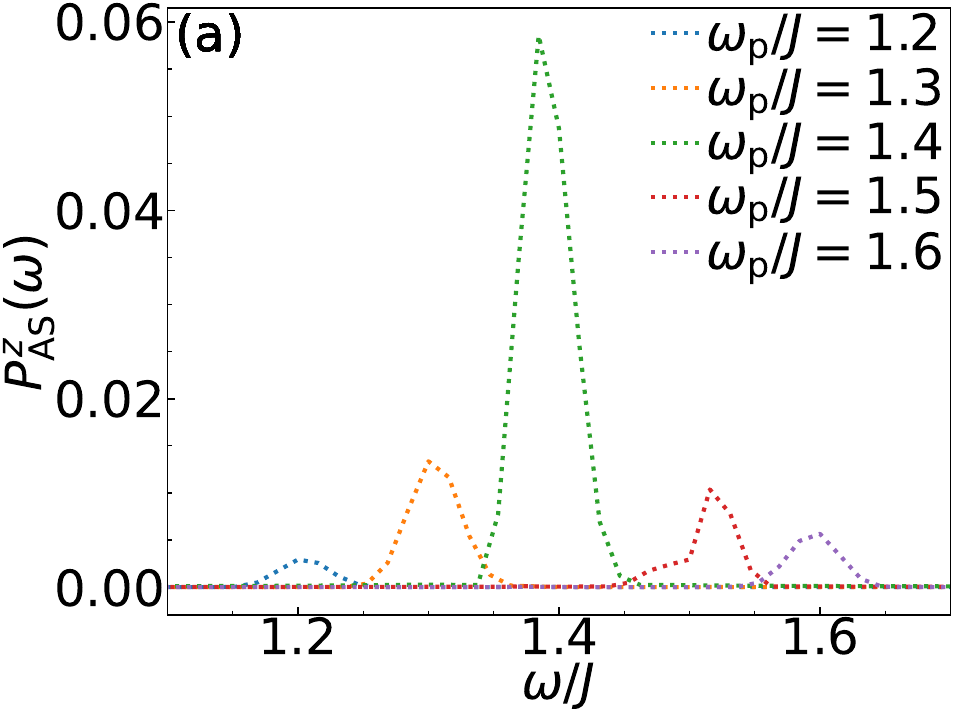}
	\end{minipage}
	\begin{minipage}[c]{0.48\hsize}
	\centering
	\includegraphics[width=\columnwidth]{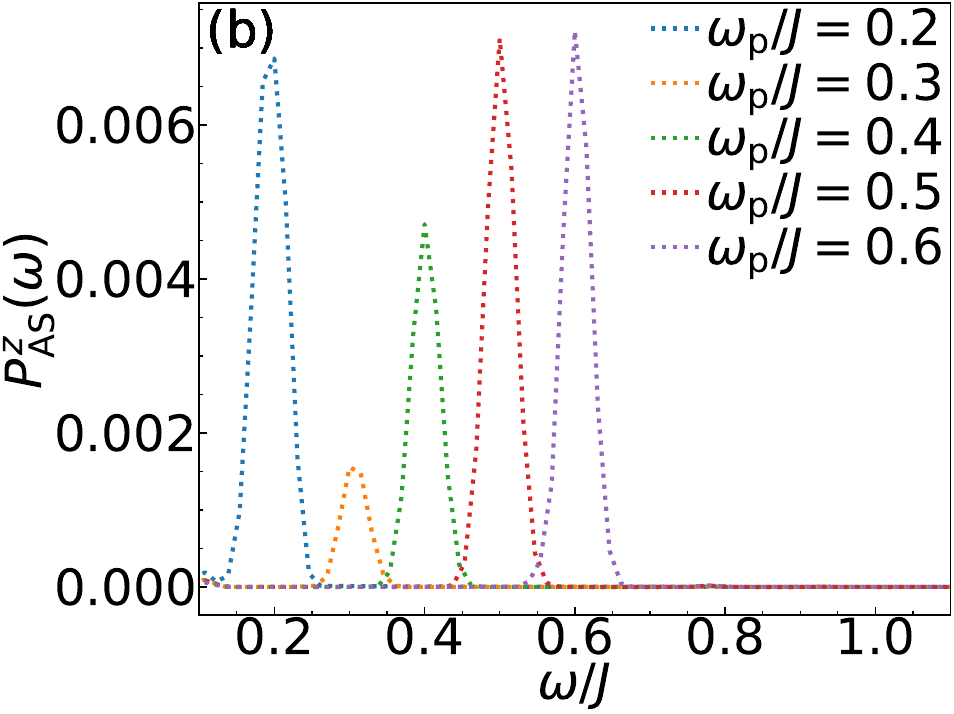}
	\end{minipage}
\end{tabular}
\caption{
$P_\mathrm{AS}^z(\omega)$ calculated in the DSP (a) under zero magnetic field and (b) under a DC magnetic field along the $z$ axis of $h_z/J=0.6$. 
We take several laser frequencies $\omega_{\rm p}$ around the twice the triplon resonance point in panel (a) and around a low-frequency region in panel (b). 
}
\label{fig:PAS_h_DSP_SS}
\end{figure}

First, we try to find the resonance frequencies of the $S=2$ modes at zero magnetic field and around the critical magnetic field of the above transition. 
%in which the PSNP is expected to appear. 
To this end, we calculate $P_\mathrm{AS}^z(\omega)$ under zero field and under a moderate DC magnetic field $h_z/J=0.6$, by applying laser pulses with different frequencies. 
The field $h_z/J=0.6$ is expected to be close to the field-induced transition point between the DSP and PSNP (i.e., the triplon-pair condensation point) in our SSM with $J'/J=0.5$. 
Figure~\ref{fig:PAS_h_DSP_SS} shows the numerical result of $P_\mathrm{AS}^z(\omega)$. 
We adopt a smaller light-matter coupling constants $|\Pi_{\rm me}E_{\rm p}|=0.01J$ and $|D_{\rm me}E_{\rm p}|=0.01J$ because we simply estimate the resonant frequencies and therefore focus only on the linear response. 
Figure~\ref{fig:PAS_h_DSP_SS} (a) depicts the spectrum around the laser frequency $\omega_{\rm p}=2\times\omega_{\rm triplon}=2\times 0.7J$ under zero magnetic field. 
The value of $\omega_{\rm triplon}=0.7J$ is the resonance frequency for single triplon as shown in the main text. 
From this figure, we find a peak at $\omega_{\rm p}/J=1.4$, which must correspond to the resonance of two triplon states. 
On the other hand, Fig.~\ref{fig:PAS_h_DSP_SS} (b) shows the linear response under $h_z/J=0.6$ and reveals that a large peak appear at $\omega_{\rm p}/J=0.2$. As we mentioned, the $(S, S^z)=(2,2)$ mode with the energy gap $\omega_{\rm p}/J=1.4$ under zero magnetic field [see Fig.~\ref{fig:PAS_h_DSP_SS} (a)] is expected to be shifted by $-2h_z/J$ resulting from the Zeeman splitting, and therefore the peak at $\omega/J=0.2$ in Fig.~\ref{fig:PAS_h_DSP_SS} (b) indeed corresponds to the $(S, S^z)=(2,2)$ mode. Figure~\ref{fig:PAS_h_DSP_SS} (b) also shows other peaks in a higher-frequency regime besides the $(S, S^z)=(2,2)$ peak, and the former peaks probably correspond to higher-spin modes with $S=3,4,\cdots$ whose excitation energies drop to a lower energy regime due to the Zeeman splitting. However, we focus on the $(S, S^z)=(2,2)$ mode at $\omega_{\rm p}/J=0.2$ because (as we mentioned) it is experimentally detectable near the phase transition between the DSP and PSNP. 
Due to finite-size effects, the excitation energy of two triplon states numerically detected here is just twice the resonance frequency of the single triplon, and it is not clear if they form bound states~\cite{Totsuka2001}. 
Nevertheless, we can conclude that at least the peaks observed in Fig.~\ref{fig:PAS_h_DSP_SS} (a) and (b) are the $S=2$ modes. 
From the discussion above, we find that the resonance frequency of the $(S,S^z)=(2,2)$ mode is $\omega_{\rm p}/J=1.4$ under zero magnetic field and it is shifted to $\omega_{\rm p}/J=0.2$ near the critical field $h_z/J=0.6$. 

%  fig. S5  %%
%%%%%%%%

Using the information about these resonance frequencies, we next argue that the HG spectrum and the dynamical symmetries in PSNP. 
Following the argument in Ref.~\cite{Batista2018}, we add a small perturbation term of the spin-nematic order $\hat{\mathscr{H}}_\mathrm{PSNP}=\lambda_\mathrm{sn}\sum\left\{\left(\hat{S}_i^+\hat{S}_j^+-\hat{S}_i^+\hat{S}_{j'}^+\right)+\mathrm{H.c.}\right\}$ into the Hamiltonian to realize the PSNP in our 16-site cluster. 
The sum of $ij$ ($ij'$) is taken over red (black) squares in Fig.~\ref{fig:EPP_SS} (a) as in the case of EPP, because the spatial ordering pattern of the PSNP is common to that of EPP. 
%The spin-nematic order breaks the U(1) spin-rotational symmetry around the $S^z$ axis. 
%From the above symmetry argument, we easily find that the dynamical symmetries of the spin-nematic phase are the same as those of EPP. 

\begin{figure}[!h]
\begin{tabular}{cc}
	\begin{minipage}[c]{0.48\hsize}
	\centering
	\includegraphics[width=\columnwidth]{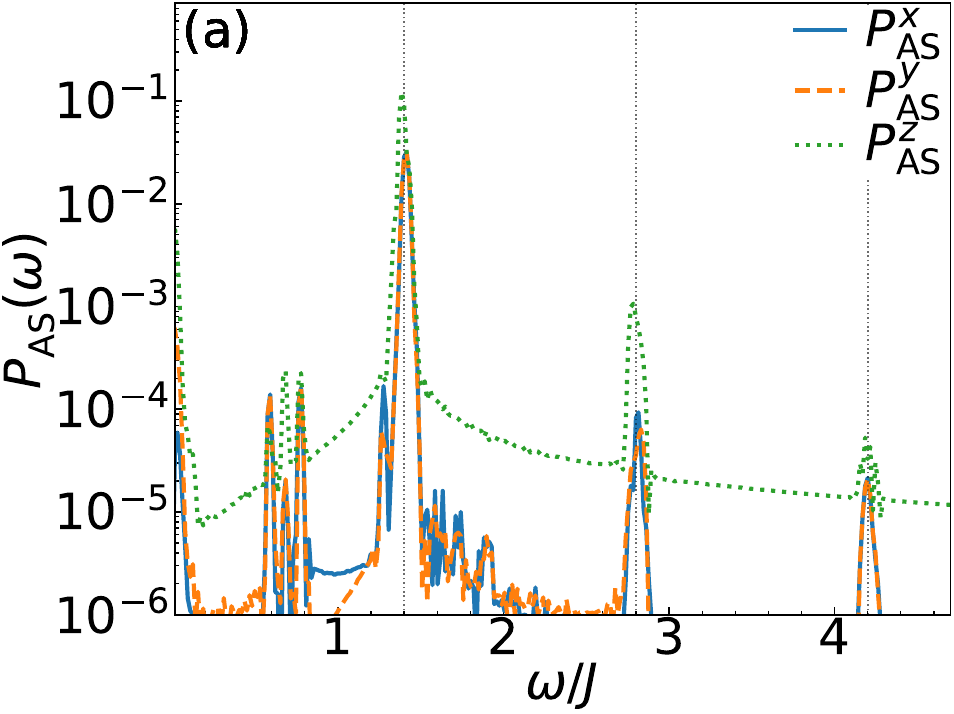}
	\end{minipage}
	\begin{minipage}[c]{0.48\hsize}
	\centering
	\includegraphics[width=\columnwidth]{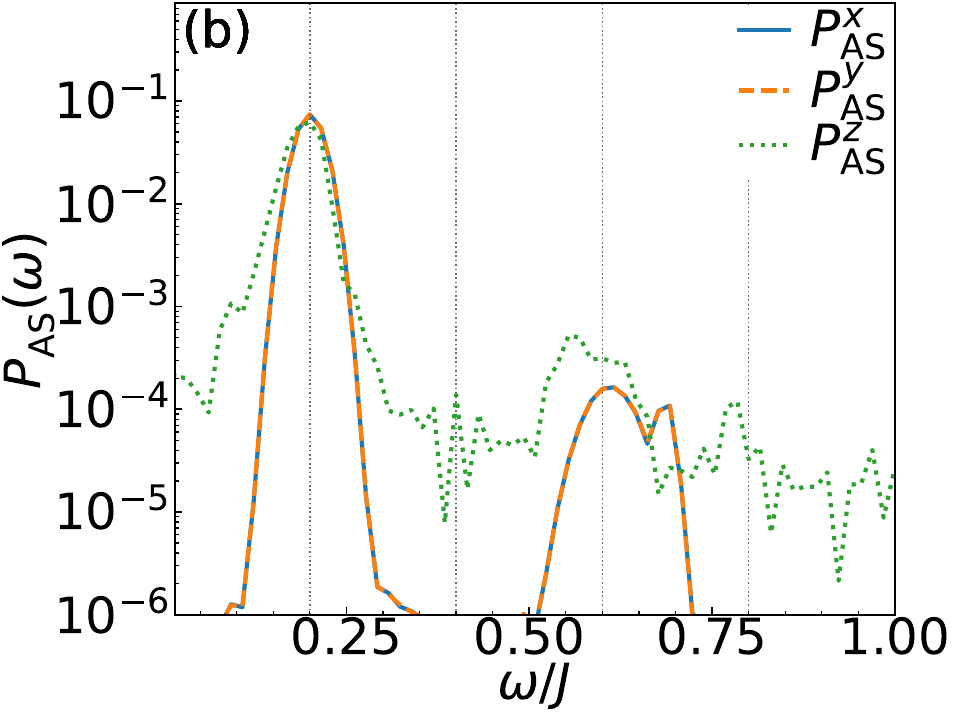}
	\end{minipage}
\end{tabular}
\caption{
HG spectra of $\bm{P}_\mathrm{AS}(\omega)$ in PSNS ($\lambda_\mathrm{sn}=0.01$) with the DC magnetic field along (a) the $x$ direction $h_x/J=0.1$ and (b) the $z$ direction $h_z/J=0.6$, respectively. 
We have used $\omega_{\rm p}/J=2\omega_{\rm triplon}=1.4$ in panel (a), while $\omega_{\rm p}/J=0.2$ in panel (b). 
Blue, orange, and green lines respectively represent the $x$, $y$, and $z$ components of $\bm{P}_\mathrm{AS}(\omega)$.
Vertical dotted lines represent integer multiples of $\omega_\mathrm{p}$. 
}
\label{fig:PAS_PSNP_SS}
\end{figure}

Below, we will consider the following two cases with spin-nematic order: The first case (i) is that the PSNP realizes in zero magnetic field, and the second case (ii) is the PSNP under a moderate DC magnetic field along the $z$ axis.
In Fig.~\ref{fig:PAS_PSNP_SS} (a), we compute $\bm{P}_\mathrm{AS}(\omega)$ in the case (i), i.e., in the spin-nematic ordered SSM with $\lambda_\mathrm{sn}=0.01$ and zero magnetic field. 
From the symmetry argument, 
the dynamical symmetry in the case (i) is shown to be the same as that of the DSP (see for more detail Table~\ref{tb:symmetry_SS}). 
Therefore, in Fig.~\ref{fig:PAS_PSNP_SS} (a), we introduce a quite weak field along the $x$ axis to distinguish the DSP and the case (i) of the PSNP. One can find from Fig.~\ref{fig:PAS_PSNP_SS} (a) that the peaks of even-order harmonics at $\omega_\mathrm{p}=2\omega_{\rm triplon}$ appear in all components of $\bm{P}_\mathrm{AS}(\omega)$ due to the magnetic field along the $x$ axis. 
On the other hand, in the DSP under a finite $h_x$, these even-order peaks appear in $P_\mathrm{AS}^{y,z}(\omega)$, whereas they disappear in 
$P_\mathrm{AS}^x(\omega)$ [see Fig.3 (b) of the main text]. 
We can also verify that even-order responses do not appear in the PSNP with $h_x=0$ like the case of DSP. 
The even-odd nature in the PSNP is the same as that of the EPP under $h_x=0$ or $h_x\neq 0$ [Fig.~\ref{fig:EPP_SS} (c)]. 
This comes from the fact that in both the PSNP and EPP, the dynamical symmetries associated with mirror operations $\sigma_{zx}$ and $\sigma_{yz}$ are violated by a transverse field $h_x$. 
Note that three triplon peaks are also seen around $\omega/J\sim 0.7$ in addition to the peaks of two triplon states at $\omega/J\sim 1.4$. 
This is because all components of $\bm{P}_\mathrm{AS}(\omega)$ are generally split by the Zeeman effect under the DC magnetic field along the $x$ axis.

Finally, we consider the case (ii) of the PSNP with a moderate magnetic field along the $z$ direction, $h_z/J=0.6$, in which the spin-nematic order more promisingly emerges in SCBO rather than zero field or a weak magnetic-field regime. 
We again add the spin-nematic perturbation with $\lambda_\mathrm{sn}=0.01$ and  choose the laser frequency to be the resonant value $\omega_{\rm p}/J=0.2$ for the $(S,S^z)=(2,2)$ mode. 
Figure~\ref{fig:PAS_PSNP_SS} (b) shows
$\bm{P}_\mathrm{AS}(\omega)$ around the frequency $\omega/J=0.2$. 
The dynamical symmetry in PNSP, which is the same as that of EPP, survives if the applied magnetic field is along the $z$ direction, any even peak does not emerge for all components of $\bm{P}_\mathrm{AS}(\omega)$. 
One finds that the fundamental and third harmonics peaks in Fig.~\ref{fig:PAS_PSNP_SS} (b) are broader than other resonant peaks in different figures. 
This would be associated to the fact that (as we mentioned) several high-spin modes gather in the low-frequency regime ($\omega/J\sim 0.2$) under the moderate field $h_z/J\sim 0.6$ [see Fig.~\ref{fig:PAS_h_DSP_SS} (b)] and these modes might contribute to the broad peaks.

%  fig. S6  %%
%%%%%%%%

%%%%%%%%%%%%%%%%%%%%%%%%%%%%%%%%%%%%
%%%%%%%%%%%%%%%%%%%%%%%%%%%%%%%%%%%%
%%%%%%%%%%%%%%%%%%%%%%%%%%%%%%%%%%%%
\subsection{Summary of selection rules}\label{sec:table_SS}
%%%%%%%%%%%%%%%%%%%%%%%%%%%%%%%%%%%%%%%
\begin{table*}[t]
\centering
\begin{tabular}{cc}
	\vspace{10pt}
	\begin{minipage}[c]{\columnwidth*2}
	\centering
	{\tabcolsep=1mm
	\begin{tabular}{c||c|c|c|c|c|c|c||c|c|c|c}
         & \multicolumn{7}{c||}{(a1)} & \multicolumn{3}{c}{(a2)}\\
         & \multicolumn{7}{c||}{dynamical symmetries in DSP} & \multicolumn{3}{c}{static symmetry}\\
	(a) symmetry operations of the U(1) breaking model &
	$P_\mathrm{S}^x$ & $P_\mathrm{S}^y$  &
	$P_\mathrm{AS}^x$ & $P_\mathrm{AS}^y$ & $P_\mathrm{AS}^z$ &
	$M^x$ & $M^z$ & FPP & EPP & PP & PSNP \\ \hline\hline
	{\color{blue}{rotation of $\pi$ around the $z$ axis ($C_2(z)\supset C_4^\pm(z)$)}} &
	$\circ$ & $\circ$ & $\circ$ & $\circ$ & $\times$ & $\circ$ & $\times$ & $\circ$ & $\circ$ & $\circ$ & $\circ$ \\ \hline
	mirror symmetry with respect to the $zx$ plane ($\sigma_{zx}=C_2(z)C_2(x)$)&
	$\times$ & $\circ$ & $\times$ & $\circ$ & $\times$ & $\circ$ & $\circ$ & $\circ$ & $\times$ & $\times$ & $\times$ \\ \hline
	{\color{red} mirror symmetry with respect to the $yz$ plane ($\sigma_{yz}=C_2(y)C_2(z)$)} &
	$\circ$ & $\times$ & $\circ$ & $\times$ & $\times$ & $\times$ & $\circ$ & $\circ$ & $\times$ & $\times$ & $\times$ \\ \hline
	{\color{blue}{rotoreflection ($S_4=C_2(x)C_2(y)C_4^+(z)$)}} &
	$\times$ & $\times$ & $\times$ & $\times$ & $\circ$ & $\times$ & $\times$ & $\times$ & $\circ$ & $\times$ & $\circ$
	\end{tabular}
	}
	\end{minipage}\\
	\begin{minipage}[c]{\columnwidth*2}
	\centering
	{\tabcolsep=1mm
	\begin{tabular}{c||c|c|c|c|c|c|c||c|c|c|c}
         & \multicolumn{7}{c||}{(b1)} & \multicolumn{3}{c}{(b2)}\\
         & \multicolumn{7}{c||}{dynamical symmetries in DSP} & \multicolumn{3}{c}{static symmetry}\\
	(b) symmetry operations of the U(1) symmetric model &
	$P_\mathrm{S}^x$ & $P_\mathrm{S}^y$  &
	$P_\mathrm{AS}^x$ & $P_\mathrm{AS}^y$ & $P_\mathrm{AS}^z$ &
	$M^x$ & $M^z$ & FPP & EPP & PP & PSNP \\ \hline\hline
        %\colorbox{yellow}{rotation of $\pi$ around the $z$ axis ($C_2(z)$)} &
	%$\circ$ & $\circ$ & $\circ$ & $\circ$ & $\times$ & $\circ$ & $\times$ & $\circ$ & $\circ$ & $\circ$ \\ \hline
	%mirror symmetry with respect to the $zx$ plane ($\sigma_{zx}=C_2(z)C_2(x)$)&
	%$\times$ & $\circ$ & $\times$ & $\circ$ & $\times$ & $\circ$ & $\circ$ & $\circ$ & $\times$ & $\times$ \\ \hline{\color{red}$\times$}\circ$ & $\circ$ \\ \hline
	{\color{green}{inversion symmetry ($I$)}} &
	$\circ$ & $\circ$ & $\circ$ & $\circ$ & $\circ$ & $\times$ & $\times$ & $\circ$ & $\circ$ & $\circ$ & $\circ$ \\ \hline
	{\color{red}rotation of $\pi$ around the $x$ axis ($C_2(x)$)} &
	$\times$ & $\circ$ & $\times$ & $\circ$ & $\circ$ & $\times$ & $\circ$ & $\circ$ & $\times$ & $\times$ & $\times$ \\ \hline
	rotation of $\pi$ around the $y$ axis ($C_2(y)$) &
	$\circ$ & $\times$ & $\circ$ & $\times$ & $\circ$ & $\circ$ & $\circ$ & $\circ$ & $\times$ & $\times$ & $\times$ \\ \hline
	{\color{blue}{rotation of $\pm\pi/2$ around the $z$ axis ($C_4^\pm(z)$)}} &
	$\times$ & $\times$ & $\times$ & $\times$ & $\times$ & $\times$ & $\times$ & $\times$ & $\circ$ & $\times$ & $\circ$
	\end{tabular}
	}
	\end{minipage}
\end{tabular}
\caption{
Column (a) [(b)] is the list of symmetry operations in the U(1)-breaking model [in the U(1)-symmetric model]. The symmetries of column (b) are broken in the U(1)-breaking SSM. 
The U(1)-breaking SSM holds finite $D'$ and $D$, while the U(1)-symmetric SSM has a finite $D'$ and $D=0$. 
%(a) List of symmetry operations existing even in the U(1) breaking case, where the intradimer DM interactions $D$ are finite, and the parameters used in the calculation are $D/J=0.025$ and $D'/J=0.015$. 
%The rotoreflection written in the bottom line means that the rotation of $\pm\pi/2$ around the $z$ axis is followed by the mirror to the $xy$ plane. 
%(b) List of symmetry operations possessed by only the U(1) symmetric case, where only the $z$ component of the interdimer DM interactions $D'$ is finite~\cite{Cepas2001}.%, and the parameters used in the calculation are $D=0$, $D'/J=0.015$. 
Both tables are separated into three parts by double vertical lines; on the left [(a) or (b)], middle  [(a1) or (b1)], and right sides [(a2) or (b2)], we write the symmetry operations $\hat{U}$ in the DSP, dynamical symmetries associated with each light-matter coupling, and static symmetries in other phases, respectively. 
Circle ($\circ$) and cross ($\times$) marks respectively mean the existence and absence of (dynamical) symmetries. 
In left panels (a) and (b), the blue text means that the symmetry is still conserved even if $\hat{\mathscr{H}}_\mathrm{Zeeman}=-\sum_ih_z\hat{S}_i^z$ is added to the SSM, while the red text means that the symmetry holds even if $\hat{\mathscr{H}}_\mathrm{Zeeman}=-\sum_ih_x\hat{S}_i^x$ is added. 
The green text means that the symmetry is not broken even if the DC magnetic field in both $x$ and $z$ directions. 
In the panels (a2) and (b2), we assume that all the ordered phases appear under zero DC magnetic field. 
Note that (as we have already discussed) the $\frac{1}2$ PP and PSNP are expected to emerge by applying a sufficient DC magnetic field along the $z$ axis $\hat{\mathscr{H}}_\mathrm{Zeeman}=-\sum_ih_z\hat{S}_i^z$, but the sign ($\circ$ or $\times$) for these two phases in (a2) and (b2) are unchanged by the DC magnetic field along the $z$ axis.
}
\label{tb:symmetry_SS}
\end{table*}
%%%%%%%%%%%%%%%%%%%%%%%%%%%%%%%%%%%%%%%%%%%%
\begin{table*}[t]
\centering
\begin{tabular}{cc}
	\vspace{10pt}
	\begin{minipage}[c]{\columnwidth*2}
	\centering
	{\tabcolsep=1mm
	\begin{tabular}{c|c}
	(a) symmetry operations of the U(1) breaking model & (a1) spin transformations\\ \hline\hline
	rotation of $\pi$ around the $z$ axis ($C_2(z)$) & $\hat{\bm{S}}_{\bm{r}}\rightarrow(-\hat{S}_{C_2(z)\bm{r}}^x,-\hat{S}_{C_2(z)\bm{r}}^y,\hat{S}_{C_2(z)\bm{r}}^z)$ \\ \hline
	mirror symmetry with respect to the $zx$ plane ($\sigma_{zx}$) & $\hat{\bm{S}}_{\bm{r}}\rightarrow(-\hat{S}_{\sigma_{zx}\bm{r}}^x,\hat{S}_{\sigma_{zx}\bm{r}}^y,-\hat{S}_{\sigma_{zx}\bm{r}}^z)$ \\ \hline
	mirror symmetry with respect to the $yz$ plane ($\sigma_{yz}$) & $\hat{\bm{S}}_{\bm{r}}\rightarrow(\hat{S}_{\sigma_{yz}\bm{r}}^x,-\hat{S}_{\sigma_{yz}\bm{r}}^y,-\hat{S}_{\sigma_{yz}\bm{r}}^z)$ \\ \hline
	rotoreflection ($S_4$) & $\hat{\bm{S}}_{\bm{r}}\rightarrow(\hat{S}_{S_4\bm{r}}^y,-\hat{S}_{S_4\bm{r}}^x,\hat{S}_{S_4\bm{r}}^z)$
	\end{tabular}
	}
	\end{minipage}\\
	\begin{minipage}[c]{\columnwidth*2}
	\centering
	{\tabcolsep=1mm
	\begin{tabular}{c|c}
	(b) symmetry operations of the U(1) symmetric model & (b1) spin transformations\\ \hline\hline
	%mirror symmetry with respect to the $xy$ plane ($\sigma_{xy}$) & $\hat{\bm{S}}_{\bm{r}}\rightarrow(-\hat{S}_{\sigma_{xy}\bm{r}}^x,-\hat{S}_{\sigma_{xy}\bm{r}}^y,\hat{S}_{\sigma_{xy}\bm{r}}^z)$ \\ \hline
	inversion symmetry (I) & $\hat{\bm{S}}_{\bm{r}}\rightarrow\hat{\bm{S}}_{I\bm{r}}$ \\ \hline
	rotation of $\pi$ around the $x$ axis ($C_2(x)$) & $\hat{\bm{S}}_{\bm{r}}\rightarrow(\hat{S}_{C_2(x)\bm{r}}^x,-\hat{S}_{C_2(x)\bm{r}}^y,-\hat{S}_{C_2(x)\bm{r}}^z)$ \\ \hline
	rotation of $\pi$ around the $y$ axis ($C_2(y)$) & $\hat{\bm{S}}_{\bm{r}}\rightarrow(-\hat{S}_{C_2(y)\bm{r}}^x,\hat{S}_{C_2(y)\bm{r}}^y,-\hat{S}_{C_2(y)\bm{r}}^z)$ \\ \hline
    rotation of $\pi/2$ around the $z$ axis ($C_4^+(z)$) & $\hat{\bm{S}}_{\bm{r}}\rightarrow(-\hat{S}_{C_4^+(z)\bm{r}}^x,\hat{S}_{C_4^+(z)\bm{r}}^y,\hat{S}_{C_4^+(z)\bm{r}}^z)$
	\end{tabular}
	}
	\end{minipage}
\end{tabular}
\caption{
Column (a) [(b)] is the list of symmetry operations in the U(1)-breaking model [in the U(1)-symmetric model]. The symmetries of column (b) are broken in the U(1)-breaking SSM. 
Column (a1) [(b1)] shows spin transformations under the symmetry operations of the left column (a) [(b)].
}
\label{tb:spin_SS}
\end{table*}

In both the main text and the SM, we have estimated the HG spectra in the SSMs under various conditions. The result has been explained by the argument based on dynamical symmetries. 
In this section, we summarize the usual static symmetries and dynamical symmetries in the SSMs. Table~\ref{tb:symmetry_SS} (a) shows these symmetries in the U(1)-breaking SSM whose Hamiltonian is given by $\hat{\mathscr{H}}=\hat{\mathscr{H}}_\mathrm{SS}+\hat{\mathscr{H}}_\mathrm{DM}$ with finite $D'$ and $D$. Table~\ref{tb:symmetry_SS} (b) shows the additional symmetries which are conserved in the U(1)-symmetric SSM with a finite $D'$ and $D=0$. 
Both tables consist of three parts; on the left [(a) or (b)], middle [(a1) or (b1)], and right sides [(a2) or (b2)], we write symmetry operations $\hat{U}$ in the DSP, dynamical symmetries associated with each light-matter coupling, and static symmetries in other phases, respectively. Circle ($\circ$) and 
cross ($\times$) marks respectively mean the existence and absence of (dynamical) symmetries. 

For example, $\circ$ mark at the position $(C_2(z), P_{\rm S}^x)$ denotes the existence of the dynamical symmetry for the spectrum $P_{\rm S}^x(\omega)$ % in the U(1)-breaking SSM 
irradiated by $\hat{\mathscr{H}}_\mathrm{ME}(t)=-E^x_{\rm p}(t)\cdot{\hat P}^x_\mathrm{S}$. 
Similarly, $\times$ mark at the position $(\sigma_{zx}, {\rm EPP})$ means the breaking of the symmetry $\sigma_{zx}$ in the EPP.% of the U(1)-breaking SSM. 

%For the symmetry operations described on the left [(a) or (b)], those with invariant $x$ ($z$) component of the spin $\hat{\bm{S}}$ are written in red text (yellow background box), and these survive symmetry even under the DC magnetic field in the $x$ ($z$) direction. 
%In the middle side [(a1) or (b1)], we note whether dynamical symmetries of the two electric polarizations ($\hat{\bm{P}}_\mathrm{S}$ and $\hat{\bm{P}}_\mathrm{AS}$) and the magnetization ($\hat{\bm{M}}$) for a linearly polarized light are present or not. 
%We denote $\circ$ when the dynamical symmetry holds and $\times$ when it does not. 
%As mentioned in the main text, under the bicircular laser with 3-fold rotational symmetry adopted in this paper, there is no dynamical symmetry at all in the SSM where this symmetry does not exist. 
%On the right side [(a2) or (b2)], we show whether the symmetry in the DSP holds for the three other phases (the FPP, EPP, and PP). 
%We write $\circ$ as the case where the symmetry holds even with the addition of the perturbation term and $\times$ as the case where the symmetry is broken by symmetry reduction. 

Blue and red texts in left panels (a) and (b) are associated with whether or not the symmetry survives in the DSP when we introduce a static Zeeman interaction $\hat{\mathscr{H}}_\mathrm{Zeeman}=-\sum_ih_\alpha\hat{S}_i^\alpha$: 
The blue text means that the symmetry is still conserved even if $\hat{\mathscr{H}}_\mathrm{Zeeman}=-\sum_ih_z\hat{S}_i^z$ is added to the SSM, 
while the red text means that the symmetry holds even if $\hat{\mathscr{H}}_\mathrm{Zeeman}=-\sum_ih_x\hat{S}_i^x$ is added. 
The green text in left panel (b) means that the symmetry is not broken even if the DC magnetic field in both $x$ and $z$ directions.

The spatial symmetry operations in Table~\ref{tb:symmetry_SS} affect not only the positions of spins but also spins themselves. 
Table~\ref{tb:spin_SS} shows how spins are transformed through those symmetry operations. 

%%%%%%%%%%%%%%%%%%%%%%%%%%%%%%%%%%%%
%%%%%%%%%%%%%%%%%%%%%%%%%%%%%%%%%%%%
%%%%%%%%%%%%%%%%%%%%%%%%%%%%%%%%%%%%
\subsection{Pulse width, system size and other parameter dependences of the line width}
\begin{figure}[t]
\vspace{0.5cm}
\begin{tabular}{cc}
	\begin{minipage}[c]{0.48\hsize}
	\centering
	\includegraphics[width=\columnwidth]{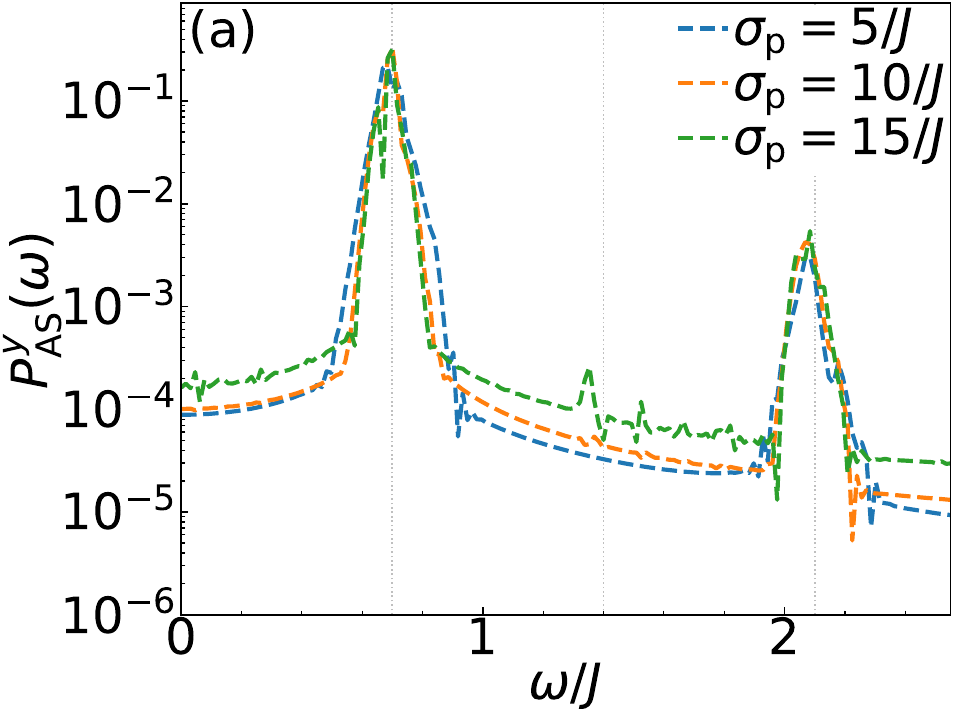}
	\end{minipage}
	\begin{minipage}[c]{0.48\hsize}
	\centering
	\includegraphics[width=\columnwidth]{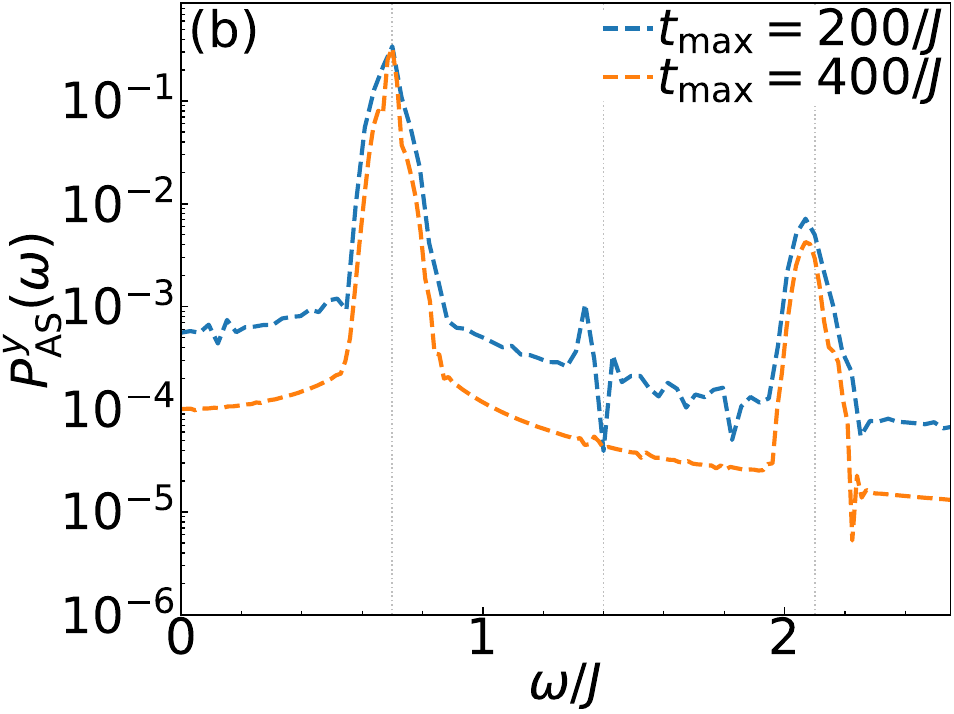}
	\end{minipage}\\
	\begin{minipage}[c]{0.48\hsize}
	\centering
	\includegraphics[width=\columnwidth]{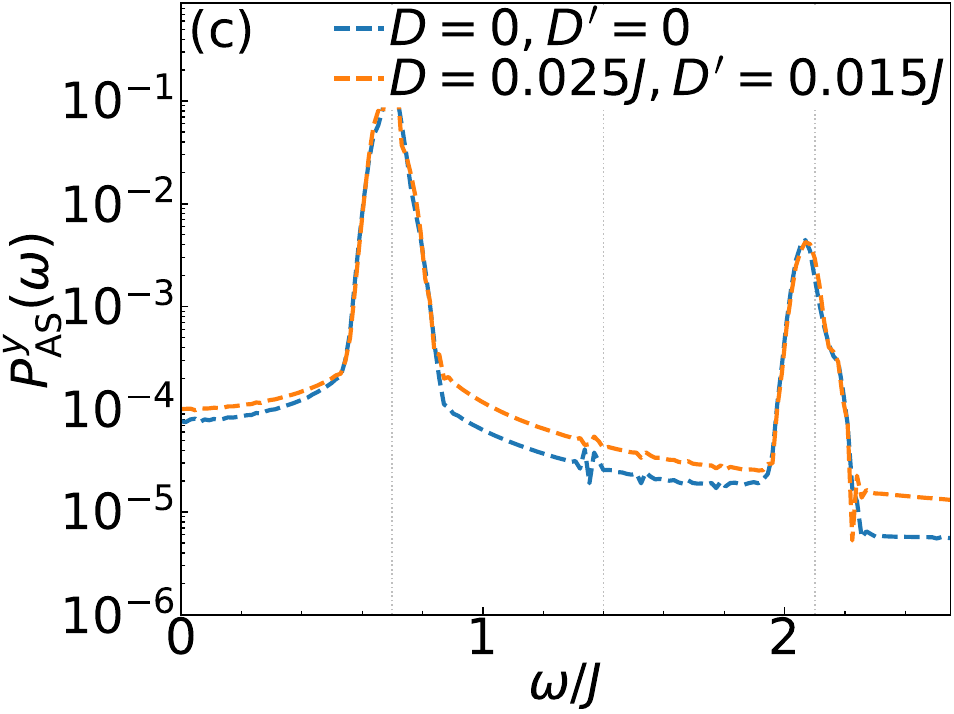}
	\end{minipage}
	\begin{minipage}[c]{0.48\hsize}
	\centering
	\includegraphics[width=\columnwidth]{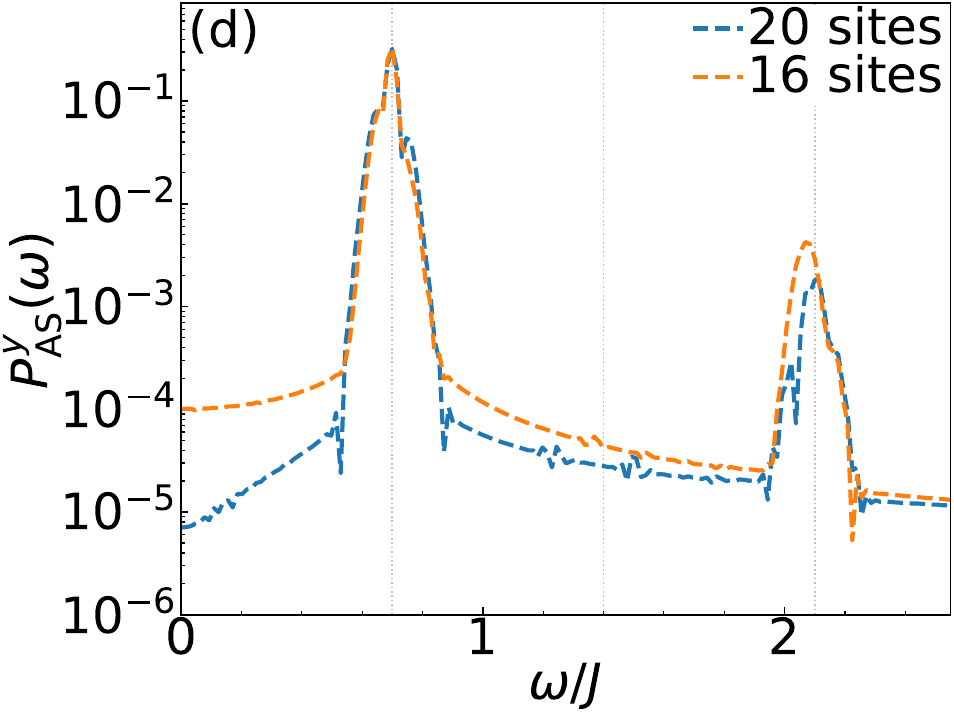}
	\end{minipage}
\end{tabular}
\caption{
HG spectra of $P_\mathrm{AS}^y(\omega)$ calculated for (a) $\sigma_\mathrm{p} J=5$, 10, and 15, (b) $t_\mathrm{max} J=200$, and 400, (c) the SU(2) symmetric case ($D=0,~D'=0$) and the U(1) breaking case with a DM interaction ($D=0.025J,~D'=0.015J$), and (d) 20- and 16- site clusters.
Other parameters are the same as those in Fig. 2 (b) of the main text. 
The orange line in each panel is identical to the line in Fig. 2 (b) of the main text. 
}
\label{fig:dependence_SS}
\end{figure}

In this section, we discuss the influence of some parameters and the system size on the resonant line width of the HG spectra. 
Firstly, we consider the dependence of the pulse width.
In our simulation, we use the pulse laser, which contains various frequencies, so that the laser and triplon (bound state) do not resonate perfectly. 
The width of the resonant peaks, therefore, depends on the pulse width. 
%and the total time range $[0, t_\mathrm{max}]$.
Figure~\ref{fig:dependence_SS} (a) depicts the $\sigma_\mathrm{p}$ dependence of the width. 
We calculate $P_\mathrm{AS}^y(\omega)$ with all parameters being the same as those in Fig. 2 (b) of the main text except for $\sigma_\mathrm{p}$. 
Blue, orange, and green lines are calculated for $\sigma_\mathrm{p} J=5, 10$ and $15$, respectively, and the orange line ($\sigma_\mathrm{p} J=10$) is identical to the line in Fig. 2 (b) of the main text. 
The larger $\sigma_\mathrm{p}$, the fewer frequencies the incident laser contains. Thus, it is expected that the width of the peak becomes narrower with increasing $\sigma_\mathrm{p}$. Figure~\ref{fig:dependence_SS} shows that the larger $\sigma_\mathrm{p}$, the narrower the width of the peak slightly, as expected. Thus, we can guess that the peaks become shaper if the applied laser pulse is closer to a continuous wave.  
We stress that the peak positions are almost unchanged, indicating that the peaks stem from the resonance of triplons. 

Secondly, we consider the dependence of the duration time $t_\mathrm{max}$. 
Our numerical calculation is done in a finite time range; therefore, the Fourier transform is not exact. 
Figure~\ref{fig:dependence_SS} (b) gives the $t_\mathrm{max}$ dependence of the HG spectra. 
We calculate $P_\mathrm{AS}^y(\omega)$ with all parameters being the same as those in Fig. 2 (b) of the main text except for $t_\mathrm{max}$. 
Blue and orange lines are calculated for $t_\mathrm{max} J=200$ and $400$, respectively, and the orange line is identical to the line in Fig. 2 (b) of the main text. 
The line width at $t_\mathrm{max}=200/J$ (blue line) is wider than that of the orange line. Namely, the line width becomes shaper with increasing $t_\mathrm{max}$. This tendency is naturally expected because the Fourier transform is close to exact form, as the duration time become longer. We again note that the peak position is unchanged and almost independent of the value of $t_\mathrm{max}$.

Thirdly, we examine the DM-term dependence of the line width. Since the DM interaction breaks the U(1) spin-rotation symmetry, the concept of pure triplon is no longer conceivable in the rigorous sense. As a result, the triplon has a finite lifetime, and its peak is expected to become broad. 
We show the dependence on DM terms in Fig.~\ref{fig:dependence_SS} (c). 
We calculate $P_\mathrm{AS}^y(\omega)$ with all parameters being the same as those in Fig. 2 (b) of the main text except for DM terms $D$ and $D'$. 
Blue and orange lines are calculated for the SU(2) symmetric case ($D=0,~D'=0$) and the U(1) breaking case ($D=0.025J,~D'=0.015J$), respectively, and the orange line is identical to the line in Fig. 2 (b) of the main text. 
Since the value of DM interactions is much smaller than the energy of unit $J$, the line width originating from the DM terms is not outstanding. 
Nevertheless, the line width of blue line is little narrower than that of the orange line, and a slight increase of the line width by DM terms is observed. 

Finally, we consider whether or not the system size of 16-site cluster we use is large enough to detect the thermodynamic result and how small the finite size effect is. 
As we briefly mentioned in the main text, 
the wavelength of the applied THz or GHz waves is much larger than the lattice space, and therefore the system is subjected to a spatially uniform electromagnetic wave. 
As a result, only modes with wave number $k_{x,y}=0$ or $\pi$ in the SSM couple with the THz or GHz waves. 
Such wave numbers are included in the 16-site system. In addition, Ref.~\cite{Miyahara2023} examining the linear responses in the SSMs, calculates the response function in larger sizes, and the width or position of the resonant peaks are shown to be almost independent of the system size. From these facts, our calculation for 16-site cluster is expected to capture the fundamental properties in the thermodynamic limit. 

We here compare the HG spectra in 16 and 20-site clusters with the other parameters being the same as those in Fig. 2 (b) of the main text. 
Figure~\ref{fig:dependence_SS} (b) shows the spectra of $P_\mathrm{AS}^y(\omega)$, where 
the blue and orange lines are calculated in 20- and 16-site clusters, respectively, and the orange line (calculated in 16 sites) is identical to the line in Fig. 2 (b) of the main text. 
Compared with these two lines, the size dependence is found to be small as expected.

From the arguments and the numerical results in this section, we can conclude that our numerical calculation in the 16-site cluster is basically reliable and detects many essential features of laser-driven dynamics in SSMs. 

%  fig. S7 %%
%%%%%%%%

%%%%%%%%%%%%%%%%%%%%%%%%%%%%%%%%%%%%
%%%%%%%%%%%%%%%%%%%%%%%%%%%%%%%%%%%%
%%%%%%%%%%%%%%%%%%%%%%%%%%%%%%%%%%%%
\subsection{Small spikes in HG spectra}
In the final section, we briefly comment on small spikes like sidebands in several HG spectra. Most of numerically computed HG spectra include small spikes in addition to dominant resonance peaks. 
These small spikes seem to be independent of $\sigma_\mathrm{p}$ and DM interactions.
Such spikes have often been observed in laser-driven HG spectra estimated by numerically solving the Schr\"odinger equation for a finite-size system. See, for example, Ref.~\cite{PhysRevLett.128.047401}, in which the laser-driven dynamics in an interacting electron model is computed by solving the Schr\"odinger equation for a finite-size system.
On the other hand, small spikes do not appear in general if we take a sufficiently long time evolution and take into account the realistic dissipation effect during the laser application. For instance, we have often obtained smooth HG spectra without weak spikes when we utilize quantum master equation to describe the time evolution of laser-driven many body systems with dissipation~\cite{Ikeda2019,Ikeda2020,Kanega2021,Kanega2024}.  

From these facts, we conclude that several spikes of the computed HG spectra in the present study are mainly attributed to the effects of the finite size, the finite time evolution and the dissipationless calculation for isolated systems. Most of them are expected to disappear in the thermodynamic limit with realistic dissipation. 

%%%%%%%%%%%%%%%%%%%%
% References
%%%%%%%%%%%%%%%%%%%%

\bibliography{supplemental_material.bbl}
%\bibliography{References}